\documentclass[twocolumn,
pra
,superscriptaddress
,floatfix
,aps
,10pt
,showpacs]{revtex4-2}

\usepackage{graphicx}
\usepackage{amssymb}
\usepackage{amsmath}
\usepackage{multirow}
\usepackage{array,tabularx}
\usepackage{xcolor}
\usepackage{comment}

\usepackage[colorlinks,urlcolor=blue,citecolor=blue,linkcolor=blue]{hyperref}

\usepackage{textcomp}

\usepackage{float}
\usepackage{upgreek}
\usepackage{makeidx}
\DeclareGraphicsExtensions{.png,.jpg,.pdf,.mps,.gif,.bmp,.eps}
\graphicspath{{figures/}}
\usepackage{physics}

\begin{document}

\title{Theory of 
quantum spin-Hall topological lasers}

\author{A. Mu$\tilde{\mathrm{n}}$oz de las Heras}
\email{Alberto.MunozHeras@uclm.es}
\affiliation{Departamento de Física, Universidad de Castilla-La Mancha, Campus Tecnológico de la Fábrica de Armas, 45004 Toledo, Spain}

\author{I. Carusotto}
\affiliation{
Pitaevskii BEC Center, CNR-INO and Dipartimento di Fisica, Universit\`a di Trento, I-38123 Trento, Italy}

\date{\today}

\begin{abstract}

We theoretically investigate a quantum spin-Hall topological laser formed by an array of dielectric ring resonators endowed of saturable gain. The system preserves time-reversal symmetry, the clockwise and counter-clockwise modes in each ring resonator acting as two pseudospin states that experience opposite synthetic magnetic fields. We consider ring resonators featuring an internal S-shaped waveguide asymmetrically coupling the two pseudospin states. In spite of the non-magnetic nature of the configuration, we show that an effective breaking of reciprocity is induced by the interplay of spatial parity breaking with saturable gain and a Kerr optical non-linearity. This enables robust single-mode topological lasing even in the presence of realistic levels of backscattering. 

\end{abstract}

\maketitle

\section{Introduction}
\label{sec:Introduction}

Since its invention in the 1960s~\cite{MAIMAN1960}, the laser has become one of the most transformative technologies in science and engineering, leading to the development of quantum optics and quantum information~\cite{Wieman1999,grynberg2010introduction} and enabling advances such as high-precision spectroscopy and metrology~\cite{Hansch1999,Ludlow2015}. The interplay between light and matter in lasing systems has inspired extensive exploration of new laser architectures, including those based on photonic crystals~\cite{Park2004}. These are periodic nanophotonic structures engineered to manipulate the propagation of light by creating photonic band gaps~\cite{Yablonovitch1987,Sajeev1987}.

Capitalizing on the ability of photonic crystals to control light through band structure engineering, the field of topological photonics has emerged as a powerful framework for designing photonic systems with robust, symmetry-protected modes that are immune to defects and disorder~\cite{Haldane2008,Lu2014,Ozawa_2019}. Inspired by the classification of topological phases in solid-state electronic  systems~\cite{Schnyder2008,Kitaev2009,Ryu_2010,Chiu2016}, topological photonic platforms implement analogues of quantum Hall and quantum spin-Hall (QSH) effects for light using structured dielectric materials, such as gyromagnetic photonic crystals~\cite{Wang_2009}, coupled ring resonators~\cite{Hafezi2011,Hafezi_2013}, bi-anisotropic metamaterials~\cite{Khanikaev2013}, or waveguide arrays~\cite{Rechtsman2013}, as well as semiconductor micropillar arrays hosting exciton–polaritons~\cite{Kartashov2017,St-Jean2017}. A key feature of these systems is the existence of unidirectional edge modes confined to the boundaries of the photonic lattice, which propagate without reflection even in the presence of imperfections. In the presence of gain, these topological modes open the door to unconventional lasing regimes where disorder resilience can be fundamentally enhanced~\cite{St-Jean2017,Bahari2017,Bandres_2018,Harari_2018,Kartashov2019,Ota_2020,Amelio_2020}.

A central challenge in designing topological lasers is to ensure that lasing occurs exclusively in the desired unidirectionally-propagating edge mode. This is specially relevant in non-magnetic systems preserving time-reversal symmetry, such as photonic QSH analogues, where edge states are typically helical~\cite{Bernevig2006,Konig2007,Hafezi2011,Hafezi_2013,Khanikaev2013} and appear in pairs of counterpropagating modes associated to opposite photonic pseudospins. This limits the laser robustness as any imperfection may couple the two pseudospin states and induce backscattering and localization effects~\cite{Ozawa_2019}. To address this issue, in Ref.~\cite{Bandres_2018} an S-shaped waveguide was introduced in each unit cell of an active ring resonator lattice: in this setup, the pseudospin corresponds to the clockwise vs. counter-clockwise propagation direction of the whispering gallery modes of each ring resonator, and the S-shaped element establishes an asymmetric coupling between the two pseudospins, selectively enhancing lasing in one pseudospin state.

Ring resonators featuring an embedded S-shaped waveguide are often known as ``Taiji'' resonators (TJRs)~\cite{Hohimer_1993,Hohimer_1993b}. The asymmetry of the coupling between counterpropagating whispering-gallery modes explicitly breaks spatial inversion $\mathcal{P}$-symmetry, and, in the presence of a photonic non-linearity, such as the saturable gain of a laser or a Kerr non-linearity (resulting in a intensity-dependent refractive index), also leads to an effective breaking of Lorentz reciprocity~\cite{MunozDeLasHeras_2021,MunozDeLasHeras_2022}.
This means that the system’s response depends on the propagation direction of the signal. However, note that the non-linear TJR preserves time-reversal symmetry.
While a single TJR has been shown to trigger unidirectional lasing in a single whispering-gallery mode~\cite{Hohimer_1993,Hohimer_1993b,MunozDeLasHeras_2021b}, the interplay of many TJRs arranged as unit cells of a QSH topological laser becomes more complex. In this many-TJR situation several questions remain unanswered, e.g., under what conditions is single-mode topological lasing possible and robust, or to what extent the S-shaped element protects topological lasing against backscattering processes coupling counterpropagating topological modes.

\begin{figure*}[t]
  \centering
  \includegraphics[width=\linewidth]{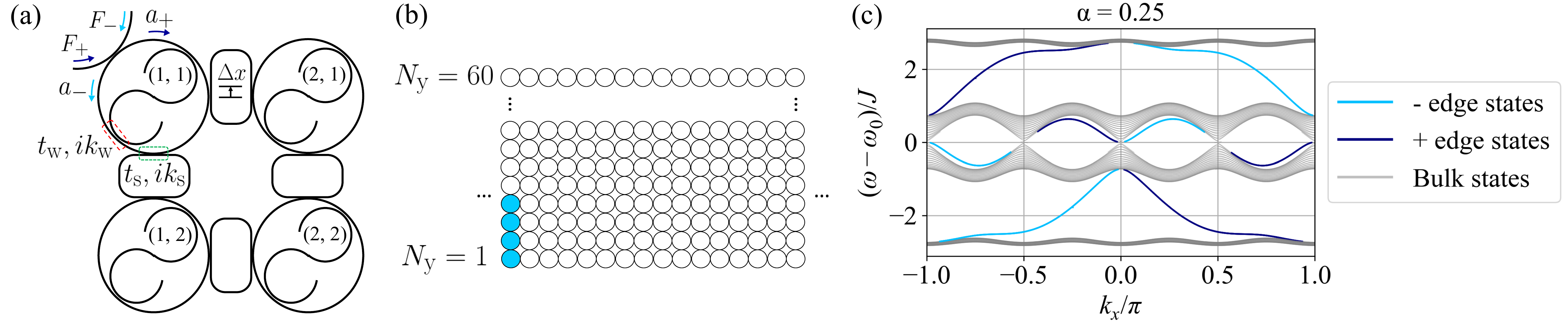}
  \caption{(a) Sketch of a $2\times 2$ example of a QSH array. For each site resonator, the field amplitude inside each counterpropagating mode is labeled $a_{\pm}$. The resonators are evanescently coupled with transmission and coupling amplitudes $t_{\rm w,s}$ and $ik_{\rm w,s}$, respectively, where the subindices w and s stand for the couplings between site and link resonators, and those between site resonators and S-shaped waveguides, respectively. $\Delta x$ is the vertical displacement introduced to generate a synthetic magnetic field for photons with opposite signs for the two $\pm$ pseudospin states. As these correspond to light propagating in the clockwise or counter-clockwise direction around each site resonator, they can be probed by means of coherent signals $F_{\pm}$ injected from either side of a bus waveguide coupled to the top left site resonator. 
  (b) Sketch of the ribbon configuration used to compute the frequency bands of a QSH insulator with flux per plaquette given by $\alpha=1/4$. Each circle represents a site resonator (link resonators are omitted for clarity). The blue circles indicate one magnetic unit cell (with size $4\times 1$ for $\alpha=1/4$). The boundary conditions are periodic in the $x$ direction and open in the $y$ direction with $N_y=60$ site resonators.
  (c) Frequency bands of the QSH insulator shown in panel (b). The vertical axis represents the frequency $\omega$ in units of the tunneling rate $J$ with respect to the resonance frequency $\omega_0$ of the site resonators. The horizontal axis corresponds to the $k_x$ wavevector, which is a good quantum number. The edge modes are plotted only for the upper edge of the lattice.
  }
  \label{fig:scheme and bands}
\end{figure*}

The goal of this work is to fill this gap by providing a comprehensive theoretical study of the physics of quantum spin Hall topological lasers based on an array of Taiji resonators. We employ coupled-mode equations to calculate the time evolution of the field amplitudes inside each resonator forming the photonic QSH array, and we compute their steady state solution. Our numerical simulations are carried out considering realistic parameters for state-of-the-art silicon photonic setups. 

We start by analyzing a passive, coherently-driven QSH topological insulator in the absence of gain, and we show that the S-shaped elements may reduce the coupling between opposite pseudospins caused by typical values of undesired backscattering present in current experiments. As a next step, we introduce a saturable gain and we study laser operation in such a photonic QSH topological array. In particular, we show that the effective breaking of reciprocity may stabilize lasing in a single topological mode. Finally, we assess how the S-shaped elements can help reducing the effect of backscattering couplings also in the active array, thus protecting lasing in a single edge mode.

The Article is organized as follows: 
In Sec.~\ref{sec:system and model} we describe the physical system (Sec.~\ref{sec:system}) and the theoretical model employed in our calculations (Sec.~\ref{sec:model}).
Sec.~\ref{sec:passive} is devoted to studying the role of TJRs in a passive QSH array.
In Sec.~\ref{sec:active} we focus on topological lasing in a TJR-based QSH array.
Conclusions are finally drawn in Sec.~\ref{sec:conclusions}. Codes to reproduce the figures of the Article are available in~\cite{github}.

\section{The physical system and the theoretical model}
\label{sec:system and model}

\subsection{The physical system}
\label{sec:system}

In this work we consider a two-dimensional (2D) square lattice of photonic ring resonators inspired by the recent experiments of Refs.~\cite{Hafezi_2013,Bandres_2018}. As an example, a $2\times 2$ version of such a lattice is sketched in Fig.~\ref{fig:scheme and bands}(a). The lattice is built from two different resonator types: site resonators, depicted in the sketch as circular, and link resonators, which are represented with a smaller, rounded square-like shape. Irrespectively of its type, each resonator hosts a pair of degenerate whispering-gallery modes propagating in opposite  directions [clockwise (CW) and counter-clockwise (CCW)]. The site resonators may feature an S-shaped waveguide inside them, which breaks spatial inversion $\mathcal{P}$-symmetry by coupling the CW mode into the CCW one, but not viceversa. In this case, we call the whole resonator array \textit{TJR lattice}. Instead, when the S-shaped elements are absent, we refer to the array as a \textit{ring resonator lattice}. 
The site resonators of an $N_{x}\times N_{y}$ lattice are labeled by the indices $(n_x,n_y)$, where $n_x=1,...,N_x$ and $n_y=1,...,N_y$. In the simulations performed in the following Sections, we take $N_x=N_y=8$. The top-left resonator $(1,1)$ is coupled to a bus waveguide that can be harnessed to probe the system through an external coherent signal.

All elements of the lattice are evanescently coupled by means of directional couplers with transmission and coupling amplitudes $t_{\rm w,s}$ and $ik_{\rm w,s}$, respectively, where the subindices w and s respectively stand for the couplings between site and link resonators, and for those between site resonators and S-shaped waveguides. As usual for lossless configurations, $t_{\rm w,s}$ and $k_{\rm w,s}$ are taken as real numbers satisfying $t^2_{\rm w,s}+k^2_{\rm w,s}=1$. 
Let then $R$ be the radius of each site resonator, and $L_{\circ}=2\pi R$ its perimeter. We can then write the perimeter of each link resonator as $L_{\rm L}=L_{\circ}+2\eta$, where the parameter $\eta$ accounts for the length difference between the two resonator types. Here we take $2\pi n_{\rm L}\eta/\lambda=\pi/2$, where $\lambda$ is the photon's wavelength and $n_{\rm L}$ is the refractive index of the waveguide. This is done to shift the link resonator modes maximally off-resonance with respect to the site resonator modes, so that a photon resonant with the site resonator modes spends much more time in the sites than in the links. The goal of link resonators is to mediate tunneling between neighboring site resonators with the desired hopping phase. 

To realize a non-trivial topology without breaking time-reversal symmetry, we follow in fact the strategy of Refs.~\cite{Hafezi2011,Hafezi_2013,Bandres_2018,Harari_2018} and introduce an upwards displacement in the link resonators connecting sites within the same row. Such a displacement $\Delta x(n_y)=\lambda\alpha (N_y-n_y)/2n_{\rm L}$ depends on the row index $n_y$ and is chosen such that photons describing CW and CCW loops around a single plaquette pick up opposite phases $\Delta\varphi=\pm 2\pi\alpha$. We can describe this mechanism in terms of effective pseudospins by associating the CW and CCW whispering gallery modes of each site resonator with two pseudospin states, denoted as $+$ and $-$, respectively. The pseudospins propagate in opposite directions and thus acquire opposite phases upon traversing the displaced link resonators, effectively experiencing opposite synthetic magnetic fields. As a result, each pseudospin realizes a time-reversed copy of a Harper-Hofstadter lattice~\cite{Harper_1955,Hofstadter1976} with flux $\pm 2\pi\alpha$ per plaquette. All together, the system ends up forming a photonic analogue of the QSH effect that preserves overall time-reversal symmetry~\cite{Hafezi2011}. Without loss of generality, in this work we take $\alpha=1/4$.

In Fig.~\ref{fig:scheme and bands}(b) we show a sketch of the ribbon configuration employed to calculate the band structure of a QSH insulator with $\alpha=1/4$. This consists of two Harper–Hofstadter lattices with opposite synthetic fluxes given by $\alpha=\pm 1/4$, corresponding to the two pseudospin components $\pm$, respectively. The system features periodic boundary conditions along the $x$ direction and open boundary conditions along $y$, so that edge modes can emerge at the upper and lower boundaries of the strip. This makes the $k_x$ wavevector a good quantum number. The magnetic unit cell comprises four site resonators along the $y$ direction, i.e., it has a $4\times 1$ shape. To calculate the band structure, we used a total of $N_y=60$ site resonators, corresponding to 15 magnetic unit cells.

As shown in Fig.~\ref{fig:scheme and bands}(c), the frequency band structure of the QSH insulator with $\alpha=1/4$ features four bulk bands. The two central bands, located within an energy scale set by the hopping energy $J$, touch each other at Dirac points. The upper and lower bands are topologically non-trivial and feature a finite Chern number $\mathcal{C} = \mp 1$ for Harper-Hofstadter lattices with $\alpha=\pm 1/4$, respectively. The Chern numbers for the Harper–Hofstadter bands may be obtained from the Diophantine equation relating them to the gap index and the flux $\alpha$~\cite{Thouless1982,Salerno2016}. The lower and upper gaps contain the four helical edge modes that propagate along the boundaries of the QSH insulator. The lowest energy gap hosts two helical modes: In one of them light propagates in the $-$ pseudospin and travels in the CW direction along the lattice boundaries (this helical mode is labelled CW-). The other helical mode, named CCW+, is formed by photons belonging to the $+$ pseudospin that propagate in the CCW direction along the array edges. Similarly, the upper topological gap hosts the helical edge modes CW+ and CCW-.

The architecture described above is readily implementable using integrated silicon photonics platforms. The necessary coupling between adjacent ring resonators can be realized through evanescent couplings between their respective waveguides.
Lattices of $\sim 10 \times 10$ site resonators, formed by individual resonators with radii on the order of $10\ \mu$m, have been demonstrated in Refs.~\cite{Hafezi_2013,Bandres_2018}. These works implemented synthetic magnetic fields using displaced link resonators, with coupling strengths, loss rates, and disorder levels fully compatible with the parameters employed in our simulations (see Sec.~\ref{sec:model}).
The S-shaped waveguide within each unit cell resonator can be fabricated using standard directional-coupler geometries~\cite{Bandres_2018,MunozDeLasHeras_2021}. Finally, optical gain can be introduced by incoherent pumping of a gain medium, such as InGaAsP quantum wells embedded in a SiO$_2$ matrix, a strategy previously employed in Ref.~\cite{Bandres_2018}.

\subsection{The coupled-mode model}
\label{sec:model}

In our numerics, we simulate the field amplitude $a^{(n_x,n_y)}_{\pm}$ in the mode of pseudospin $\pm$ of each site resonator $(n_x,n_y)$ by means of a temporal coupled-mode theory. The corresponding intensity at each site resonator is given by $|a^{(n_x,n_y)}_{\pm}|^2$. Considering a coherent signal with amplitude $F_{\pm} = \tilde{F}_{\pm}e^{-i\omega t}$ and frequency $\omega$ driving the the top left site resonator of the lattice, we can rewrite the field amplitudes in each resonator as $a^{(n_x,n_y)}_{\pm}(t)=\tilde{a}^{(n_x,n_y)}_{\pm}(t)\,e^{-i\omega t}$. 

The time-dependent coupled-mode equations for the slow variables $\tilde{a}^{(n_x,n_y)}_{\pm}(t)$ read
\begin{widetext}
\begin{align}
    i\frac{d\tilde{a}^{\rm (n_x,n_y)}_{\pm}}{dt}
    &=
    (\omega_0-\omega)\tilde{a}^{(n_x,n_y)}_{\pm}
    -\frac{n_{\rm NL}}{n_{\rm L}}\omega_0\left[|\tilde{a}^{(n_x,n_y)}_{\pm}|^2+2|\tilde{a}^{(n_x,n_y)}_{\mp}|^2\right]\tilde{a}^{(n_x,n_y)}_{\pm}
    +i\frac{P^{(n_x,n_y)}}{1+\frac{1}{n_{\rm s}}\left[|\tilde{a}^{(n_x,n_y)}_{\pm}|^2+2|\tilde{a}^{(n_x,n_y)}_{\mp}|^2\right]}\tilde{a}^{(n_x,n_y)}_{\pm}
    \nonumber\\
    &
    + J\left[ 
    \tilde{a}^{(n_x,n_y-1)}_{\pm}
    +\tilde{a}^{(n_x,n_y+1)}_{\pm} 
    +e^{\mp i 2\pi \alpha (N_y-n_y)} \tilde{a}^{(n_x-1,n_y)}_{\pm} 
    +e^{\pm i 2\pi \alpha (N_y-n_y)} \tilde{a}^{(n_x+1,n_y)}_{\pm} 
    \right]
    \nonumber\\
    &
    -i\gamma_{\rm T}
    \tilde{a}^{(n_x,n_y)}_{\pm}
    -i\gamma^{(n_x,n_y)}_{\rm out}
    \tilde{a}^{(n_x,n_y)}_{\pm}
    +\beta_{\pm,\mp}\tilde{a}^{(n_x,n_y)}_{\mp}
    +\beta^{(n_x,n_y)}_{\rm in}\tilde{F}_{\pm}.
\label{eq:coupled mode}
\end{align}
\end{widetext}
A step-by-step derivation of the equations above can be found in Appendix~\ref{app:coupled mode}. Note that Eq.~\eqref{eq:coupled mode}, employed to obtain the results presented in the Main Text of this work, considers local optical non-linearities for both the Kerr and the gain saturation contributions. An analog analysis for non-local non-linearities (yet restricted within each site resonator) can be found in Appendix~\ref{app:nonlocal nonlinearity gain}. 

We denote by $\omega_0$ the resonance frequency of the site resonators. A $\chi^{(3)}$ optical Kerr non-linearity~\cite{butcher_cotter_1990} can be taken into account by means of the non-linear refractive index $n_{\rm NL}$. Here, $n_{\rm L}$ stands for the linear refractive index. A saturable gain can be added by considering a finite pump rate $P^{\rm (n_x,n_y)}$ at the desired sites $(n_x,n_y)$, with  saturation intensity $n_{\rm s}$. It is important to notice that Eq.~\eqref{eq:coupled mode} describes a class-A laser in which the dynamics of the reservoir triggering lasing is assumed to be much faster than any other timescale, and therefore it has been integrated out in the saturable gain term~\cite{Amelio_2020,Secli2019,Loirette2021}.

The link resonators have been integrated out in the coupled-mode equations and their role is accounted by the effective couplings between site resonators, with tunneling amplitude $J = c k^2_{\rm w} / {2L_\circ n_{\rm L}}$. 
The synthetic magnetic field is described by the $y$-dependence of the tunneling phase along the $x$ direction, corresponding to a magnetic flux $\pm 2\pi\alpha$ per lattice plaquette for each pseudospin.

The single resonator total loss rate $\gamma_{\rm T}=\gamma_{\rm A}+\gamma_{\rm s}$ is the sum of the absorption losses $\gamma_{\rm A}$ and the loss rate due to the coupling with the S element $\gamma_{\rm s}=ck^2_{\rm s}/L_{\circ}n_{\rm L}$. 
The additional loss rate $\gamma^{\rm (n_x,n_y)}_{\rm out}=\delta_{n_x,1}\delta_{n_y,1}ck^2_{\rm w}/2L_{\rm s}n_{\rm L}$ is due to the radiative coupling of the top left (1,1) resonator to the bus waveguide. Note that, since the link resonators are anti-resonant with the site resonators, for frequencies $\omega$ in the neighborhood of $\omega_{0}$ losses introduced by the coupling between site and link resonators are exponentially suppressed.

The parameters $\beta_{\pm,\mp}=\beta^{\rm (S)}_{\pm,\mp}+\beta^{\rm (BS)}_{\pm,\mp}$ describe a coupling between opposite pseudospins, either due to the S-shaped element ($\beta^{\rm (S)}_{\pm,\mp}$) or due to backscattering ($\beta^{\rm (BS)}_{\pm,\mp}$). We consider an S-shaped waveguide with radius $R/2$ which couples the $+$ into the $-$ pseudospin, leading to the coupling rates $\beta^{\rm (S)}_{+,-}=0$ and $\beta^{\rm (S)}_{-,+}=-i2ck^{2}_{\rm s}e^{i\omega_{0} n_{\rm L}L_{\circ}/2c}e^{i\omega_{0} n_{\rm L}L_{\circ}/4c}/L_{\circ}n_{\rm L}$. 
In this work, when a finite backscattering is included we assume it to be Hermitian, i.e., satisfying $\beta^{\rm (BS)}_{\mp}={\beta^{\rm (BS)}}^{*}_{\pm}$. This represents a symmetric,
conservative exchange of energy between opposite pseudospins.
Finally, the strength of the coherent coupling to incident radiation in the bus waveguide is given by $ \beta^{(n_x,n_y)}_{\rm in}=-\delta_{n_x,1}\delta_{n_y,1} c k_{\rm w}/L_{\circ}n_{\rm L}$.

In our simulations, we choose realistic parameters for our model based on a silicon photonics implementation, such as those of Ref.~\cite{Hafezi_2013,Bandres_2018}. In particular, we take a resonance frequency $\omega_0=2\pi c/1.55$ $\mu$m, a linear refractive index $n_{\rm L}=3.5$, a site-resonator radius $R=20$ $\mu$m, a loss rate $\gamma_{\rm A}=2.73$ GHz, and a coupling coefficient $k_{\rm w}=0.31$, which leads to a tunneling rate $J=33$ GHz. These values have been employed throughout the present work.

In the following Sections we apply the model in Eq.~\eqref{eq:coupled mode} to describe the physics of both passive and active photonic QSH insulators, focusing on the effect of backscattering and the role of the S-shaped element. 
We are specifically interested in the long-time stationary states $\tilde{a}^{\rm (n_x,n_y)}_{\pm}(t\rightarrow\infty)$ of the field amplitudes, which we obtain by solving Eq.~\eqref{eq:coupled mode} with a 4th order Runge-Kutta algorithm.

\section{Coherently-driven, passive spin-Hall topological insulator}
\label{sec:passive}

\begin{figure}[!t]
  \centering
  \includegraphics[width=\linewidth]{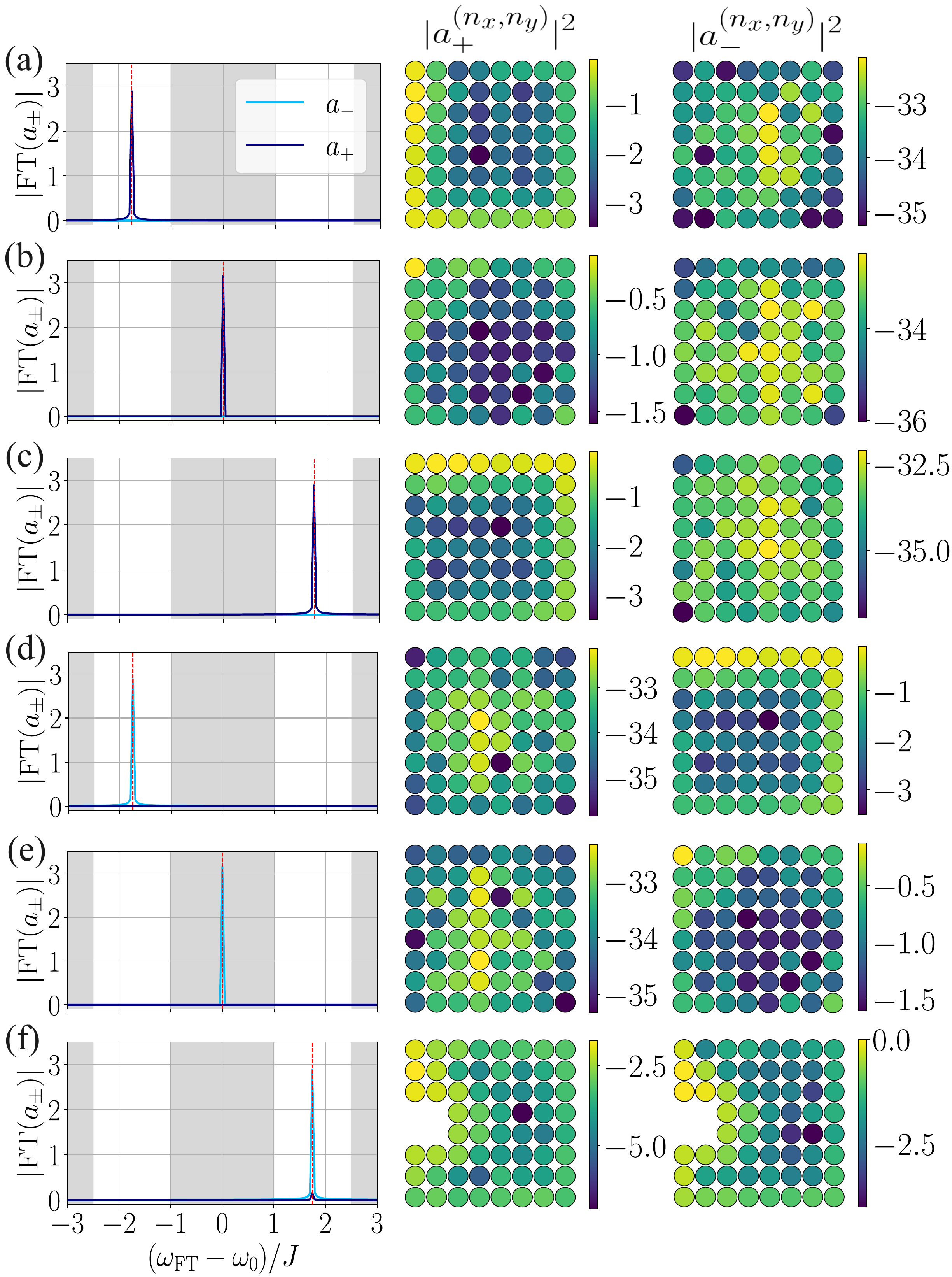}
  \caption{Coherently driven passive QSH insulator formed by an array of ring resonators without S-shaped elements. Left panels show the temporal Fourier transform $|$FT$(a_{\pm})|$ of the fields on the top-left resonator as a function of the frequency $\omega_{\rm FT}$. 
  Central (right) panels show the intensity $| a^{(n_x,n_y)}_{\pm}|^2$ in the $+$ ($-$) pseudospin at each site resonator (normalized by the maximum intensity in the lattice and in logarithmic scale). The small residual intensity visible in the non-pumped pseudospin originates solely from numerical noise. 
  In panels (a-e) no backscattering is present and
  (a) $\omega=\omega_0-1.75J$, $F_+=1$, and $F_-=0$;
  (b) $\omega=\omega_0$, $F_+=1$, and $F_-=0$;
  (c) $\omega=\omega_0+1.75J$, $F_+=1$, and $F_-=0$;
  (d) $\omega=\omega_0-1.75J$, $F_+=0$, and $F_-=1$;
  (e) $\omega=\omega_0$, $F_+=0$, and $F_-=1$.
  Panel (f) has $\omega=\omega_0+1.75J$, $F_+=0$ and $F_-=1$, and includes a Hermitian backscattering $|\beta^{\rm (BS)}_{\pm,\mp}|=0.04J$ with a random phase at each site resonator; furthermore a few sites are removed around the left edge.}
  \label{fig:passive}
\end{figure}

\begin{figure}[t]
  \centering
  \includegraphics[width=\linewidth]{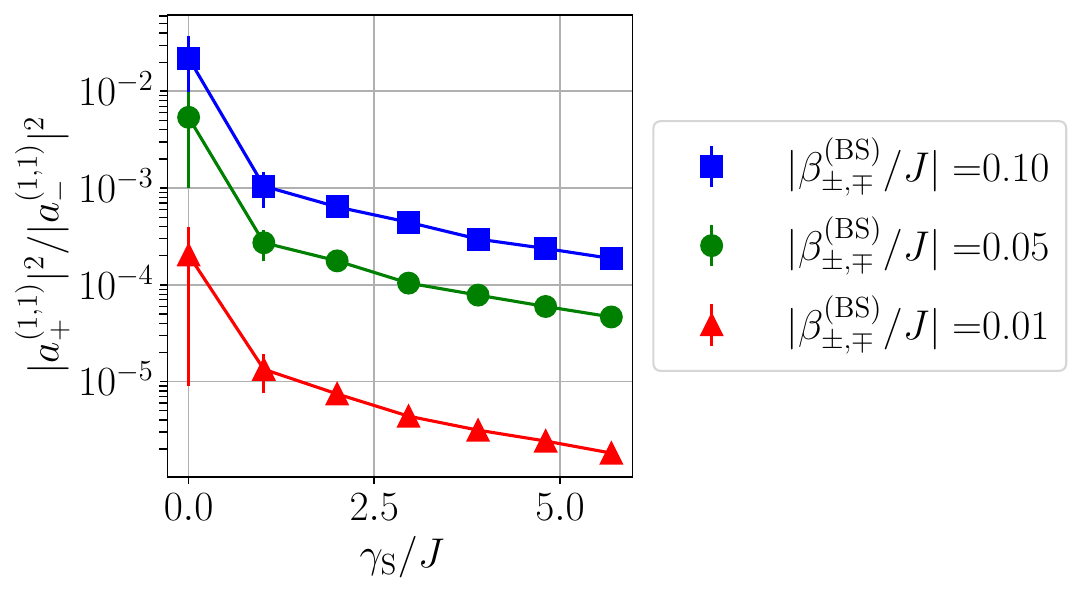}
  \caption{Passive QSH insulator formed by an array of  TJRs in the presence of Hermitian backscattering. The array is coherently driven through the bus waveguide by means of a signal coupling to the CCW- edge mode. We plot the ratio $|a^{(1,1)}_+|^2/|a^{(1,1)}_-|^2$ between the intensities in the $+$ and $-$ pseudospin of the top-left $(1,1)$ resonator as a function of the S-waveguide coupling loss rate $\gamma_{\rm s} = ck^2_{\rm s} / L_\circ n_{\rm L}$ for three values of the backscattering modulus $|\beta^{\rm (BS)}_{\pm,\mp}|$ (in units of the tunneling rate $J$). Squares, circles, and triangles are calculated as the average over 10 realizations of the system. The backscattering phase for each site resonator is chosen randomly in every realization. Error bars represent the standard deviation. Lines are added as a guide to the eye.}
  \label{fig:passive BS}
\end{figure}

In this Section we consider a coherently driven, passive photonic analog of a QSH insulator in the absence of gain. In our coupled-mode formalism of Eq.~\eqref{eq:coupled mode}, this amounts to setting $P^{(n_x,n_y)}=0$ for all $n_x$ and $n_y$ in the 2D array. We also restrict to the linear system case, i.e., we set $n_{\rm NL}=0$ throughout this Section.

We probe the system with a monochromatic coherent signal of frequency $\omega$ traveling through the bus waveguide from either side, thus coupling to either the + or - pseudospin. 
After obtaining the steady-state of the coupled-mode equations~\eqref{eq:coupled mode}, we perform a Fourier transform (FT) of the field amplitudes $a_{\pm}$ in the top left resonator (1,1). The results for a ring resonator lattice  without S-shaped elements (i.e., $\beta^{\rm (S)}_{\pm,\mp}=0$) in the absence of backscattering ($\beta^{\rm (BS)}_{\pm,\mp}=0$) are shown in Fig.~\ref{fig:passive}(a-e), where we plot the frequency spectrum $|$FT$(a_{\pm})|$ in the top-left resonator as a function of the FT frequency $\omega_{\rm FT}$ for several values of $\omega$ and for the two possible propagation directions, each one exciting one of the pseudospins: it is clear from the plots on the left column that photons inside the lattice feature the same frequency as the coherent signal driving them. Moreover, only the pseudospin coupled to the driving signal is excited.
Although this was expected, the results of Fig.~\ref{fig:passive} allow us to verify the consistency of our numerical model and to provide a reference for comparison with the situation in which backscattering or a saturable gain are present in the array. 
For each case, on the central and right columns we plot, respectively, the intensity $| a^{(n_x,n_y)}_{\pm}|^2$ in pseudospin $+$ and $-$ at each site resonator of the lattice.

When the coherent drive frequency lies inside of a topological gap, one of the helical edge states is excited. The calculations are in agreement with the frequency bands shown in Fig.~\ref{fig:scheme and bands}(b): $\omega=\omega_0-1.75J$, $F_+=1$, and $F_-=0$ excites the CCW+ helical mode [Fig.~\ref{fig:passive}(a)];
$\omega=\omega_0+1.75J$, $F_+=1$, and $F_-=0$ excites the CW+ helical mode [Fig.~\ref{fig:passive}(c)]; 
$\omega=\omega_0-1.75J$, $F_+=0$, and $F_-=1$ excites the CW- helical mode [Fig.~\ref{fig:passive}(d)]; and $\omega=\omega_0+1.75J$, $F_+=0$, and $F_-=1$ excites the CCW- helical mode [Fig.~\ref{fig:passive}(f)]. In contrast, when the drive frequency overlaps with the bulk bands or is at the Dirac point, light may penetrate the bulk  [Fig.~\ref{fig:passive}(b,e)].
Although inside of each topological gap two edge modes are available for each frequency $\omega$, in the absence of backscattering and of the S-shaped waveguide coupling, the dynamics of each pseudospin remains independent. Consequently, only the edge mode corresponding to the driven pseudospin is excited, while the opposite pseudospin state remains unpopulated.

To prove the topological protection of the helical edge modes, in Fig.~\ref{fig:passive}(f) we include in the ring resonator array a  Hermitian backscattering of fixed modulus $|\beta^{\rm (BS)}_{\pm,\mp}|=0.04J$ (similar to the one present in state-of-the-art experiments such as the one of Ref.~\cite{Hafezi_2013}) and random, independent phases for each site resonator. We also remove a $2\times 2$ plaquette of resonators on left edge of the lattice. For the chosen parameters, the coherent drive excites the CCW- edge mode. While the backscattering introduces a finite coupling between the two pseudospins (as can be demonstrated by the small $+$ pseudospin component present in the frequency spectrum), the edge state bypasses the defect without entering in the bulk. This is the smoking gun of topological robustness.

As a next step, we consider a QSH insulator with TJRs as site resonators, and we study the role of the S-shaped element in the presence of Hermitian backscattering. We coherently drive the lattice by means of a signal propagating through the bus waveguide and coupling to the CCW- edge mode [i.e., at a frequency $\omega=\omega_0$+1.75J, as shown in Fig.~\ref{fig:passive}(f)]. In Fig.~\ref{fig:passive BS} we plot the ratio between the intensities in the $+$ and $-$ pseudospins of the top left $(1,1)$ resonator $|a^{(1,1)}_+|^2/|a^{(1,1)}_-|^2$ as a function of the S-waveguide coupling loss rate $\gamma_{\rm s}$ for three values of the backscattering modulus $|\beta^{\rm (BS)}_{\pm,\mp}|$. Each point is calculated by averaging 10 realizations of the QSH insulator. In each of them, the backscattering phases are randomly and independently chosen at each site resonator, and we solve the coupled-mode equations~\eqref{eq:coupled mode} to find the steady state. 

Our results show that the intensity in the $+$ pseudospin decreases with increasing $\gamma_{\rm s}$. For the three values of backscattering analyzed, TJRs with $\gamma_{\rm s}=5.7J$ lead to a ratio $|a^{(1,1)}_+|^2/|a^{(1,1)}_-|^2$ approximately two orders of magnitude smaller than the analog ring-resonator QSH array. This effect is due to the increased value of the total resonator losses $\gamma_{\rm T}=\gamma_{\rm A}+\gamma_{\rm s}$ that includes a contribution from the S-shaped element, in agreement with Eq.~(23) of Ref.~\cite{MunozDeLasHeras_2021b}.
In particular, for values of backscattering similar to those present in Ref.~\cite{Hafezi_2013} (i.e., $|\beta^{\rm (BS)}_{\pm,\mp}|=0.05J$), $\gamma_{\rm s}=5.7J$ leads to an intensity in the $+$ pseudospin around four orders of magnitude smaller than that in the $-$ pseudospin.

\section{Spin-Hall topological laser}
\label{sec:active}

In this Section, we consider topological laser operation when gain is included on the edge resonators of the photonic QSH insulator. In our model, this is described by introducing a finite saturable gain with a pump rate $P^{(n_x,n_y)}=P$ at resonators with $n_x=1,N_x$ and $n_y=1,N_y$. On the other hand, bulk resonators feature $P^{(n_x,n_y)}=0$. We label $N_{\rm P}=2(N_x+N_y-2)$ the number of pumped resonators.
We remind the reader that the results of this Section correspond to an $8\times8$ QSH array. This size was chosen to balance numerical tractability with experimentally realistic device dimensions. Although we expect our conclusions to hold for larger arrays, in the case of smaller arrays finite-size effects can be important. Topological lasing in smaller QSH arrays is studied in Appendix~\ref{app:effect system size}.

\begin{figure}[t]
  \centering
  \includegraphics[width=\linewidth]{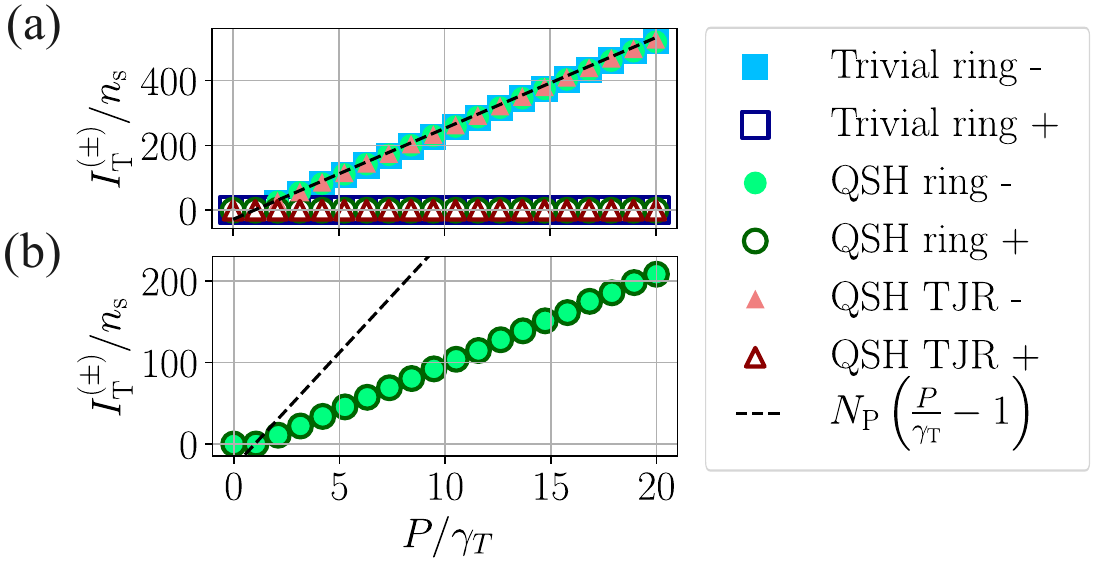}
  \caption{(a) Total intensity $I^{(\pm)}_{\rm T}$ in the $\pm$ pseudospin state 
  as a function of the pump rate $P$. 
  Data are calculated for a trivial array of ring resonators (blue squares), a QSH array of ring resonators (green circles), and a QSH lattice with TJRs asymmetrically coupling the $+$ into the $-$ pseudospin (red triangles) with S-waveguide coupling loss rate $\gamma_{\rm s}=0.2J$. Solid (empty) markers correspond to the $-$ ($+$) pseudospin. No backscattering is considered. Each data set corresponds to a single realization starting from random initial conditions for the field amplitudes. The dashed black line is the expected dependence of the total intensity $I^{(-)}_{\rm T}/n_{\rm s}=N_{\rm P}(P/\gamma_{\rm T}-1)$ for $N_{\rm P}$ independent pumped resonators. (b) Laser operation in a state with equal intensities in the two pseudospins. The total intensity in each pseudospin $I^{(\pm)}_{\rm T}$ is plotted as a function of the pump rate $P$. In this case, a Hermitian backscattering of magnitude $|\beta^{\rm (BS)}_{\pm,\mp}|=0.05J$ and random phases at each site resonator is considered.}
  \label{fig:active single realization}
\end{figure}

\begin{figure*}[htbp]
  \centering
  \includegraphics[width=\linewidth]{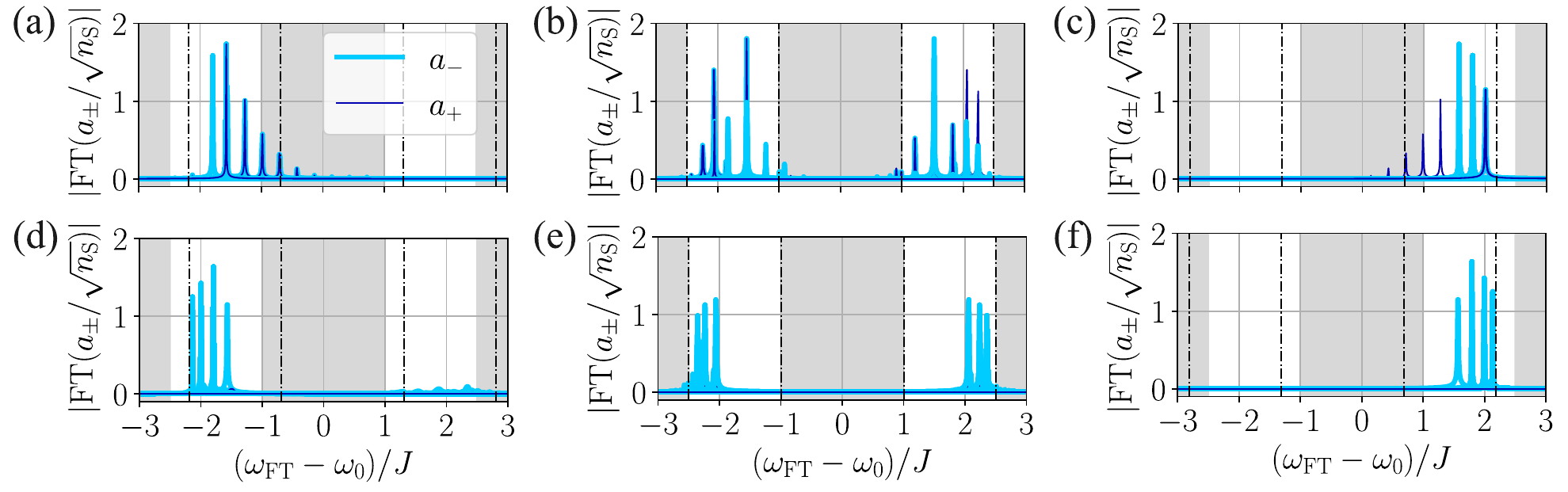}
  \caption{Frequency-dependent spectrum $|$FT$(a_{\pm})|$ of the field amplitudes $a_{\pm}$ 
  in the top-left (1,1) resonator of an active QSH lattice, as a function of the Fourier transform frequency $\omega_{\rm FT}$. 
  The pump rate is fixed to $P=4\gamma_{\rm T}$. No backscattering is considered (i.e., $\beta^{\rm (BS)}_{\rm \pm,\mp}=0$). Data is calculated using 40 realizations, each starting from random initial conditions. 
  Dark (light) blue solid lines correspond to the $+$ ($-$) pseudospin. The shaded regions represent the bulk bands of the QSH array in the linear regime. (a-c) QSH laser of ring resonators with different values of the Kerr non-linearity: (a) $n_{\rm NL}=-10^{-5}n^{-1}_{\rm s}$, (b) $n_{\rm NL}=0$, and (c) $n_{\rm NL}= 10^{-5}n^{-1}_{\rm s}$. (d-f) QSH laser of TJRs with $\gamma_{\rm s}=0.2J$ and different values of the Kerr non-linearity: (d) $n_{\rm NL}=-10^{-5}n^{-1}_{\rm s}$, (e) $n_{\rm NL}=0$, and (f) $n_{\rm NL}= 10^{-5}n^{-1}_{\rm s}$. In all panels, dashed-dotted lines represent the shift of the topological bandgaps due to the Kerr non-linearity.}
  \label{fig:active fourier spectrum}
\end{figure*}

We start by analyzing the response of the system in a single realization as the pump rate $P$ increases. To do this, in Fig.~\ref{fig:active single realization}(a) we plot the total intensities in each pseudospin $I^{(\pm)}_{\rm T}=\sum^{N_x}_{n_x=1}\sum^{N_y}_{n_y=1}|a^{(n_x,n_y)}_\pm|^2$ as a function of $P$. We compare the results obtained for a trivial array of ring resonators (i.e., with $\alpha=0$ and $\gamma_{\rm s}=0$), a QSH array with ring resonators (i.e., featuring $\alpha=0.25$ and $\gamma_{\rm s}=0$), and a QSH lattice with TJRs coupling the $+$ into the $-$ pseudospin (with $\alpha=0.25$ and $\gamma_{\rm s}=0.2J$). In these calculations, we do not include any Kerr non-linearity ($n_{\rm NL}=0$) and we assume no backscattering ($\beta^{\rm (BS)}_{\pm,\mp}=0$). 

As expected, for $P<\gamma_{\rm T}$ the system displays the trivial solution $I^{(\pm)}_{\rm T}=0$ in the three cases. However, at $P=\gamma_{\rm T}$ a Hopf bifurcation occurs and the onset of lasing takes place~\cite{erneux2010laser}. To facilitate the benchmarking, we restrict our comparison to realizations in which lasing is triggered in the $-$ pseudospin. In the three cases, for $P>\gamma_{\rm T}$ single-pseudospin lasing in the $-$ pseudospin is observed with a total intensity $I^{(-)}_{\rm T}$ linearly dependent on $P$. The values of $I^{(-)}_{\rm T}$ reached by the three laser types are slightly below the expected total intensity for $N_{\rm P}$ independent pumped resonators, which would follow $I^{(-)}_{\rm T}=N_{\rm P}(P/\gamma_{\rm T}-1)$~\cite{MunozDeLasHeras_2021b}. This is mostly due to the finite lateral extension of the edge mode that slightly penetrates inside the non-pumped bulk. 
The slope of the $I^{(-)}_{\rm T}$ vs $P$ lines for $P>\gamma_{\rm T}$ is anyway very similar for the three studied laser types.

While in the TJR-based QSH array the only solution possible for $P>\gamma_{\rm T}$ is (i) single-pseudospin lasing in the $-$ pseudospin, for the trivial and QSH lasers of ring resonators there are two other types of possible solutions: 
(ii) Single-pseudospin lasing in the $+$ pseudospin. (iii) Lasing in the two pseudopsins with equal intensities [shown in Fig.~\ref{fig:active single realization}(b) for the ring-resonator QSH array].
The unidirectionality of lasing in cases (i) and (ii) is stabilized by the mode-competition effect mediated by the saturable gain term of the coupled-mode equations~\eqref{eq:coupled mode}. When one of the whispering-gallery modes of the site resonators [in the case of realizations shown in Fig.~\ref{fig:active single realization}(a) it is the $-$ pseudospin] starts with a slightly higher intensity, this in turn inhibits amplification in the opposite pseudospin. Over time, this imbalance grows and steers the system towards a unidirectional lasing regime, where only the favored pseudospin survives, while intensity in the other is extinguished~\cite{MunozDeLasHeras_2021b}. However, for ring resonators with no S-shaped element, Eq.~\eqref{eq:coupled mode} also admits solutions with the same intensity in the $+$ and $-$ pseudospins. Although in this case a small initial power imbalance could rapidly drive the system towards a single-pseudospin lasing solution, our numerics indicate that solutions of type (iii) survive even in the presence of a Hermitian backscattering with magnitude $|\beta^{\rm (BS)}_{\pm,\mp}|=0.05J$ (similar to that present in the setup of Ref.~\cite{Hafezi_2013}) and random, independent complex phases in each site of the lattice. 

As a next step, in Fig.~\ref{fig:active fourier spectrum}(a-c) we plot the Fourier transform of the field amplitudes $|$FT$(a_{\pm})|$ in the top-left (1,1) resonator, for 40 realizations of the QSH ring-resonator laser. Each realization is computed independently, considering random values of the initial field amplitudes. No backscattering is considered, i.e., we take $\beta^{\rm (BS)}_{\pm,\mp}=0$. To address the role of the Kerr non-linearity, we consider three values of the non-linear refractive index: $n_{\rm NL}=- 10^{-5}n^{-1}_{\rm s}$ [panel (a)], $n_{\rm NL}=0$ [panel (b)], and $n_{\rm NL}= 10^{-5}n^{-1}_{\rm s}$ [panel (c)]. Note that the position in frequency of the topological bandgaps will be shifted due to the presence of a finite Kerr non-linearity. This shift can be estimated from Eq.~\eqref{eq:coupled mode} as $-n_{\rm NL}/n_{\rm L}(P/\gamma_{\rm T}-1)$. As the S-waveguide coupling is set to zero, the system randomly chooses to lase either solely into the $+$ or $-$ pseudospin, or with equal intensities into the two pseudospins at the same time. 

As shown in panels (a-c), lasing is always triggered at frequencies lying in the topological gaps. However, the sign of the Kerr non-linearity controls the topological gap in which lasing takes place. To understand this, note that Ref.~\cite{Loirette2021} performed a Bogoliubov dynamical stability analysis of the small fluctuations around the steady-state of a topological laser based on a bosonic Harper-Hofstadter model (i.e., analog to consider a single pseudospin in the QSH model employed in our work). There, it was  found that, when the curvature of the edge modes dispersion has the same sign as the non-linear refractive index $n_{\rm NL}$, the system is dynamically unstable. This result agrees with our observations [see Fig.~\ref{fig:scheme and bands}(b)]: for $n_{\rm NL}<0$ lasing is triggered in the bottom topological gap, in which the dispersion curvature is positive, for $n_{\rm NL}=0$ lasing can take place in either one of the two topological gaps, and for $n_{\rm NL}>0$ lasing can only occur in the upper topological gap, in which the dispersion curvature is negative. However, even in the presence of a finite Kerr non-linearity, lasing is still possible in either one of the two pseudospins. This is a manifestation of the fact that Lorentz reciprocity is preserved when one employs ring resonators as the unit cell of the QSH laser.

On the other hand, when one considers a lattice of TJRs, the combination of the $\mathcal{P}$-inversion symmetry introduced by the S-shaped element of the TJRs and the optical non-linearity provided by the saturable gain (and, additionally, the Kerr non-linearity) may lead to an effective breaking of reciprocity~\cite{Potton_2004}.  
A consequence of this in our case is that lasing in the TJR-based photonic QSH insulator is only possible in the topological modes associated to one of the pseudospins (in our case, the $-$ one). Fig.~\ref{fig:active fourier spectrum}(d-f) displays an analog analysis to that carried in panels (a-c) of the same figure, but in this case for a QSH laser with TJRs featuring $\gamma_{\rm s}=0.2J$. As explained in the previous paragraph, the sign of the Kerr non-linearity also controls the topological gap in which lasing takes place in this case: the bottom one for $n_{\rm NL}<0$, the two ones for $n_{\rm NL}=0$, and the upper one for $n_{\rm NL}>0$. However, differently from the QSH laser of ring resonators, when a finite Kerr non-linearity is introduced in the QSH insulator of TJRs, the asymmetry introduced by the S-shaped waveguide leads to  topological lasing in a single topological gap and in a single pseudospin state. 
The dynamical stability of such steady-state solutions is confirmed in Appendix~\ref{app:dynamical stability}.

It is important to note that the spectra shown in Fig.~\ref{fig:active fourier spectrum} are calculated from many different realizations of laser operation starting from different initial conditions. Each peak corresponds to lasing into a different super-mode of the giant effective ring resonator formed by the perimeter of the whole lattice device: for each realization, a single such super-mode is selected and the emission is fully and stably single-mode.

\begin{figure}[htbp]
  \centering
  \includegraphics[width=\linewidth]{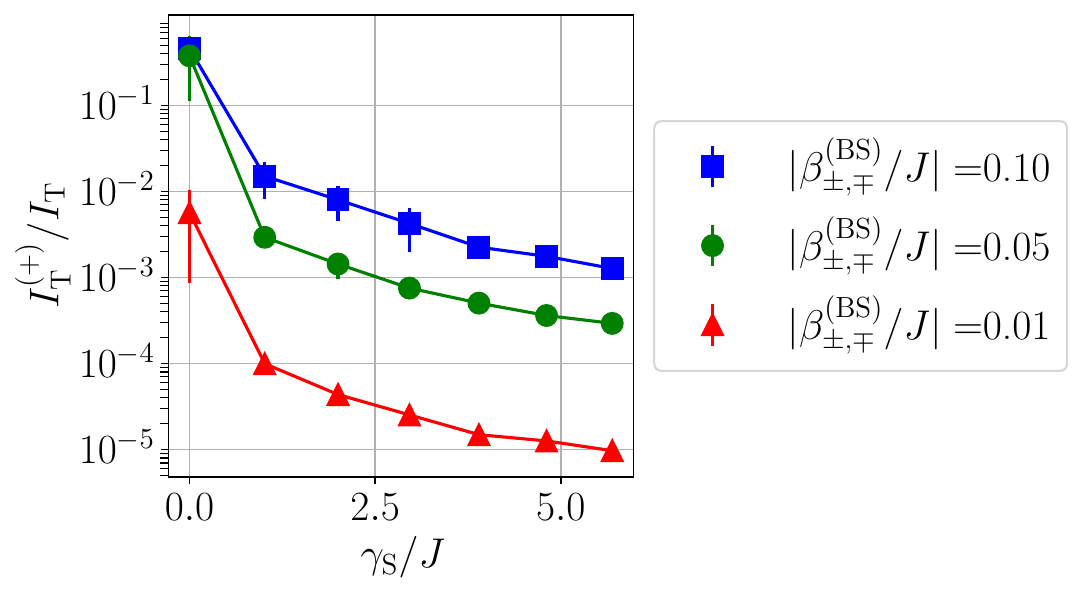}
  \caption{Active QSH lattice of TJRs with Hermitian backscattering. The pump rate is fixed to $P=4\gamma_{\rm T}$, and the non-linear refractive index is set to $n_{\rm NL}= 10^{-5}n^{-1}_{\rm s}$. The TJRs couple the $+$ into the $-$ pseudospin. We plot the total intensity in the $+$ pseudospin $I^{(+)}_{\rm T}$ in units of the total intensity in the two pseudospins $I_{\rm T}$ as a function of the S-waveguide coupling loss rate $\gamma_{\rm s}=ck^2_{\rm s}/L_\circ n_{\rm L}$ for three values of the Hermitian backscattering strength $|\beta^{\rm (BS)}_{\pm,\mp}|$. 
  Squares, circles, and triangles are calculated as the average over 10 realizations of the system. In each realization the backscattering phases and the initial values of the field amplitudes are randomly chosen for each site. Error bars represent the standard deviation. Lines are added as a guide to the eye.}
  \label{fig:active BS}
\end{figure}

Finally, we consider a finite backscattering and address the role of the S-shaped element in the TJR-based QSH laser. As a figure of merit we choose the ratio between the total intensity in the $+$ pseudospin $I^{(+)}_{\rm T}=\sum^{N_x}_{n_x=1}\sum^{N_y}_{n_y=1}|a^{(n_x,n_y)}_{+}|^2$ and the total intensity in the two pseudospins $I_{\rm T}$. We plot this quantity as a function of the S-waveguide coupling loss rate $\gamma_{\rm s}$ in Fig.~\ref{fig:active BS}, for several values of the Hermitian backscattering coupling $\beta^{\rm (BS)}_{\pm,\mp}$. We fix the pump rate to $P=4\gamma_{\rm T}$, and we set the non-linear refractive index to $n_{\rm NL}= 10^{-5}n^{-1}_{\rm s}$. Similarly to Fig.~\ref{fig:passive BS}, each data point is obtained using 10 realizations of the QSH laser, starting from random initial conditions and considering random, independent phases for the backscattering coupling at each site resonator of the lattice. 

The results show that topological lasing in a single pseudospin state is robust up to realistic values of the backscattering $|\beta^{\rm (BS)}_{\pm,\mp}|=0.05J$ similar to the experiment of Ref.~\cite{Hafezi_2013}. Furthermore, our calculations show that the presence of the S-waveguide significantly reduces the intensity in the $+$ pseudospin, with around two orders of magnitude of difference between $\gamma_{\rm s}=5.7J$ and $\gamma_{\rm s}=0$, for all values of $|\beta^{\rm (BS)}_{\pm,\mp}|$ considered. Once again, in agreement with Eq.~(23) of Ref.~\cite{MunozDeLasHeras_2021b}, a main origin of the suppressed backscattering stems from the increased value of $\gamma_{\rm T}$ when the S-shaped element is introduced.
In particular, for the value $|\beta^{\rm (BS)}_{\pm,\mp}|=0.05J$ inspired by Ref.~\cite{Hafezi_2013}, the use of TJRs with $\gamma_{\rm s}=5.7J$ reduces the ratio $I^{(+)}_{\rm T}/I_{\rm T}$ to almost $10^{-4}$.

\section{Conclusions}
\label{sec:conclusions}

In this work we provided a comprehensive study on the use of Taiji resonators (TJRs) in quantum spin-Hall (QSH) topological lasers. Such resonators feature an asymmetrical coupling between the two photonic pseudospins of the QSH array which results in laser operation in a single pseudospin state. 

In particular, we focused on a silicon photonics implementation of the QSH insulator with $\alpha = 1/4$, which features two topological gaps. Each gap hosts a pair of counter-propagating helical edge modes associated with opposite photonic pseudospins. Altogether, the system supports four such modes that propagate along the lattice edges with well-defined chiralities. In contrast to QSH arrays of ring resonators, where Lorentz reciprocity is preserved and lasing is possible in the two pseudospins, we show that the effective breaking of reciprocity induced by the interplay of spatial parity breaking and saturable gain in the TJRs triggers deterministic single-pseudospin lasing. Furthermore, we show that Kerr non-linearities can be harnessed to further restrict the laser emission to a single topological gap. Altogether, this leads to strict single-mode topological lasing. We finally show that such laser operation is robust against backscattering effects coupling opposite pseudospin states up to realistic strengths of the backscattering coupling. 

Our results confirm the promise of  photonic quantum spin-Hall arrays of Taiji resonators as a promising route to robust single-mode topological lasing. Natural future directions of research include the exploration of alternative non-linearities for enhanced mode control~\cite{Wan2019,Lukin2023,Zhu2025}, and the study of non-reciprocal elements as building blocks of higher-order topological photonic systems~\cite{Schindler2018,Zhu2021,Wu2023}.

\acknowledgements

AMH acknowledges support from Fundación General CSIC's ComFuturo program, which has received funding from the European Union's Horizon 2020 research and innovation program under the Marie Skłodowska-Curie grant agreement No. 101034263. IC acknowledges support from the Provincia Autonoma di Trento; from the Q@TN Initiative; from the National Quantum Science and Technology Institute through the PNRR MUR Project PE0000023-NQSTI, co-funded by the European Union - NextGeneration EU.

\appendix

\section{Derivation of the coupled-mode equations}
\label{app:coupled mode}

\setcounter{figure}{0}
\renewcommand{\thefigure}{A\arabic{figure}}

\begin{figure}[t]
    \centering
    \includegraphics[width=0.5\textwidth]{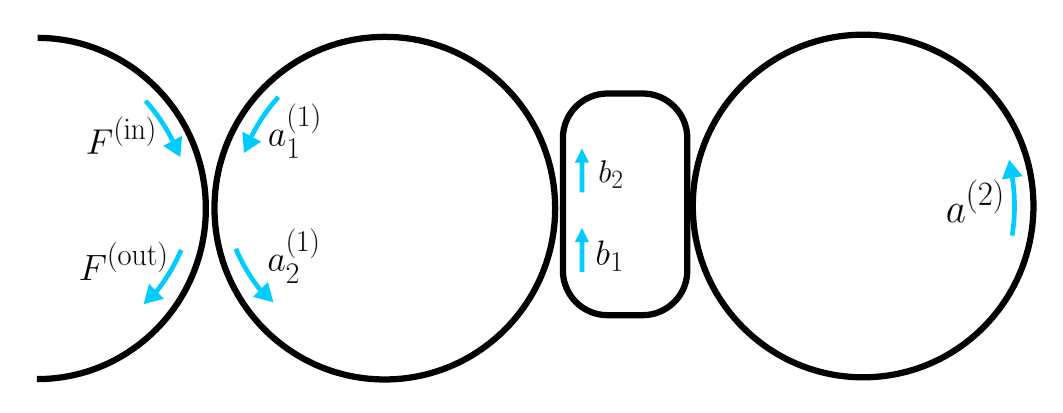}
    \caption{Sketch of the top left $2\times 1$ plaquette of site resonators (with perimeter $L_{\circ}$) of the photonic quantum QSH insulator connected through a link resonator (with perimeter $L_{\rm L}$). The left site resonator is coupled to a bus waveguide. The arrows indicate the propagation direction of the fields.}
    \label{fig:SpinHall_1x2}
\end{figure}

In this Appendix we explicitly derive the coupled-mode equations~\eqref{eq:coupled mode} for the field amplitudes inside each site resonator of the photonic analog of the QSH insulator. For analytical simplicity, we restrict to the simplest case of a $2 \times 1$ lattice, as sketched in Fig.~\ref{fig:SpinHall_1x2}. Two site resonators of perimeter $L_{\circ}$ are coupled via a racetrack-shaped resonator with perimeter $L_{\rm L}$. The left site resonator is also coupled to a bus waveguide. For simplicity, for the moment we do not take into account the S-shaped elements, the saturable gain, the Kerr non-linearity, or internal losses. We also assume that the position of the link resonator is not shifted with respect to the central horizontal axis of the plaquette.

When the plaquette is probed by a signal with field amplitude $F^{\rm (in)}$, the $-$ pseudospin is excited. We label by $a^{(1)}_{1,2}$ the field amplitudes of the $-$ pseudospin in the left site resonator before and after the coupling with the bus waveguide. In a similar manner, $b_{1,2}$ are the field amplitudes in the link resonator before and after the coupling with the left site resonator. Finally, $a^{(2)}$ is the field amplitude in the right site resonator at an opposite position in the circumference to the coupling with the link resonator. The output amplitude through the bus waveguide is labelled $F^{\rm (out)}$. We name the transmission and coupling amplitudes between the bus waveguide and the left site resonator $t_{\rm w}$ and $k_{\rm w}$. The transmission and coupling amplitudes between the link resonator and the site resonators are labelled $t_{\rm \ell}$ and $k_{\rm \ell}$. In the two cases, the amplitudes are taken as real numbers satisfying $t^2_{\rm w,\ell} + k^2_{\rm w,\ell}=1$.

We now write down the system of six equations relating the six unknown field amplitudes and the inhomogeneous term given by the input field:
\begin{align}
    a^{(1)}_{1} & =a^{(1)}_{2}e^{i\frac{\omega}{c}n_{\rm L}\frac{L_{\circ}}{2}} t_{\rm w} e^{i\frac{\omega}{c}n_{\rm L}\frac{L_{\circ}}{2}} + ik_{\rm \ell} b_{1}e^{i\frac{\omega}{c}n_{\rm L}\frac{L_{\circ}}{2}} \label{eq:CMSH_1} \\
    a^{(1)}_{2} & =t_{\rm w}a^{(1)}_{1} + ik_{\rm w}F^{\rm (in)} \label{eq:CMSH_2} \\
    b_{1} & =b_{2}e^{i\frac{\omega}{c}n_{\rm L}\frac{L_{\rm L}}{2}} t_{\rm \ell} e^{i\frac{\omega}{c}n_{\rm L}\frac{L_{\rm L}}{2}} 
    \nonumber\\&
    + ik_{\rm \ell}a^{(2)}e^{i\frac{\omega}{c}n_{\rm L}\frac{L_{\circ}}{2}}e^{i\frac{\omega}{c}n_{\rm L}\frac{L_{\rm L}}{2}} \label{eq:CMSH_3} \\
    b_{2} & =t_{\rm \ell} b_{1}+ik_{\rm \ell}e^{i\frac{\omega}{c}n_{\rm L}\frac{L_{\circ}}{2}}a^{(1)}_{2} \label{eq:CMSH_4} \\
    a^{(2)} & = e^{i\frac{\omega}{c}n_{\rm L}\frac{L_{\circ}}{2}} t_{\rm \ell} e^{i\frac{\omega}{c}n_{\rm L}\frac{L_{\circ}}{2}}a^{(2)} 
    \nonumber\\&
    + ik_{\rm \ell}b_{2}e^{i\frac{\omega}{c}n_{\rm L}\frac{L_{\rm L}}{2}} e^{i\frac{\omega}{c}n_{\rm L}\frac{L_{\circ}}{2}} \label{eq:CMSH_5} \\
    F^{\rm (out)} & = t_{\rm w}F^{\rm (in)} + ik_{\rm w}a^{(1)}_{1}.  \label{eq:CMSH_6} 
\end{align}

Our aim is to end with two of equations describing the field amplitudes inside the site resonators $a^{(1)}_{1}$ and $a^{(2)}$ (which we will promote to dynamical variables) as a function of all the other field amplitudes. We first combine Eqs.~(\ref{eq:CMSH_2}-\ref{eq:CMSH_5}) to integrate out the fields in the link resonator. After a bit of manipulation, we arrive to the equation for $a^{(2)}$:
\begin{align}
    a^{(2)}e^{-i\frac{\omega}{c}n_{\rm L}L_{\circ}}
    &=
    t_{\rm \ell}\left(1-k^2_{\rm \ell}\frac{e^{i\frac{\omega}{c}n_{\rm L}L_{\rm L}     }}{1-t^2_{\rm \ell}e^{i\frac{\omega}{c}n_{\rm L}L_{\rm L}     }  }\right)a^{(2)} \nonumber \\
    &-k^2_{\rm \ell}\left(1+t^2_{\rm \ell}\frac{e^{i\frac{\omega}{c}n_{\rm L}L_{\rm L}     }}{1-t^2_{\rm \ell}e^{i\frac{\omega}{c}n_{\rm L}L_{\rm L}     }  }\right) e^{i\frac{\omega}{c}n_{\rm L}\frac{L_{\rm L}}{2}} 
    \nonumber\\
    &\times
    \left( t_{\rm w} a^{(1)}_{1} +ik_{\rm w}F^{\rm (in)} \right).
\label{eq:a2_1_notime}
\end{align}
Since we know that the resonance frequency of the site resonators $\omega_0$ satisfies 
\begin{align}
    e^{i\frac{\omega_0}{c}n_{\rm L}L_{\circ}  }=1,
\end{align}
it is legitimate to multiply by the complex phase $ \exp(i\omega_0n_{\rm L}L_{\circ}/c  )$ in the left-hand side of Eq.~\eqref{eq:a2_1_notime}. Assuming that $\omega$ is always in the vicinity of the resonance frequency, we can Taylor expand the left-hand side, obtaining
\begin{align}
    a^{(2)}\left( 1-i\frac{\omega}{c}n_{\rm L} L_{\circ}+i\frac{\omega_0}{c}n_{\rm L}L_{\circ} \right).
\end{align}
At this point we can change from reciprocal to real space by employing the Fourier transform
\begin{align}
    -i\omega a^{(2)}(\omega) \rightarrow \frac{d}{dt}a^{(2)}(t).
\end{align}
By keeping terms up to quadratic order in the couplings $k_{\rm w,s}$ we finally arrive to the coupled-mode equation for the right site resonator:
\begin{align}
    i\dot{a}^{(2)}&=\omega_0 a^{(2)}
    -i\frac{c}{n_{\rm L}L_{\circ}}k^2_{\rm \ell}\frac{e^{i\frac{\omega}{c}n_{\rm L}\frac{L_{\rm L}}{2}  }  }{1-e^{i\frac{\omega}{c}n_{\rm L}L_{\rm L}  }  }a^{(1)}_{1}
    \nonumber\\&
    -i\frac{c}{n_{\rm L}L_{\circ}}\frac{k^2_{\rm \ell}}{2}\frac{1+e^{i\frac{\omega}{c}n_{\rm L}L_{\rm L}  }  }{1-e^{i\frac{\omega}{c}n_{\rm L}L_{\rm L}  }  }a^{(2)}
    ,
\label{eq:CoupledModeRight}
\end{align}
where we can identify the second term in the right-hand side of the equation with the incoming light due to the coupling with the left site resonator mediated through the link resonator. The last term on the second line of the equation has a negligible importance in the case under consideration of link resonators anti-resonant with the site resonators, where $e^{i\frac{\omega_0}{c}n_{\rm L}L_{\rm L} } \simeq -1$. 
The main advantage of this equation is that we got rid of the field amplitudes in the link resonator. The role of the link resonators is in fact implicitly accounted by the losses and coupling terms.

Following a similar procedure, we employ Eqs.~(\ref{eq:CMSH_1}-\ref{eq:CMSH_4}) to derive an expression for the field amplitude in the left site resonator $a^{(1)}_{1}$, i.e.,
\begin{align}
    a^{(1)}_{1}e^{-i\frac{\omega}{c}n_{\rm L}L_{\circ} }
    &= t^2_{\rm w}a^{(1)}_{1} +ik_{\rm w}t_{\rm w}F^{\rm (in)} 
    \nonumber\\&
    -k^2_{\rm \ell}t_{\ell}\frac{e^{i\frac{\omega}{c}n_{\rm L}L_{\rm L} } }{1-t^2_{\rm \ell}e^{i\frac{\omega}{c}n_{\rm L}L_{\rm L} } } \left( t_{\rm w}a^{(1)}_1 + ik_{\rm w}F^{\rm (in)} \right)
    \nonumber\\&
    - k^2_{\rm \ell}\frac{e^{i\frac{\omega}{c}n_{\rm L}\frac{L_{\rm L}}{2} } }{1-t^2_{\rm \ell}e^{i\frac{\omega}{c}n_{\rm L}L_{\rm L} } } a^{(2)}.
\end{align}
By changing to real space and keeping terms up to second order in the couplings $k_{\rm w,s}$, as we did for the right site resonator, we finally arrive to the temporal coupled-mode equation
\begin{align}
    i\dot{a}^{(1)}_{1}&=\omega_0 a^{(1)}_{1}-\frac{c}{n_{\rm L}L_{\circ}}k_{\rm w}F^{\rm (in)}   
    \nonumber\\&
     -i\frac{c}{n_{\rm L}L_{\circ} }\frac{k^2_{\rm w}}{2}a^{(1)}_{1}
    -i\frac{c}{n_{\rm L}L_{\circ}}k^2_{\rm \ell}\frac{e^{i\frac{\omega}{c}n_{\rm L}\frac{L_{\rm L}}{2} } }{1-e^{i\frac{\omega}{c}n_{\rm L}L_{\rm L} } }a^{(2)} 
    \nonumber\\&
    -i\frac{c}{2n_{\rm L}L_{\circ} }\frac{k^2_{\rm w}+\left(2 k^2_{\rm \ell}-k^2_{\rm w}\right)e^{i\frac{\omega}{c}n_{\rm L}L_{\rm L} } }{1-e^{i\frac{\omega}{c}n_{\rm L}L_{\rm L} } }a^{(1)}_{1}.
\label{eq:CoupledModeLeft}
\end{align}
On the right-hand side of this equation, the second to fourth terms represent the light injected from the bus waveguide, radiative losses into the bus waveguide, and the light coupled into the left site resonator from the right site resonator. As before, the last term has a negligible importance when the link resonators are anti-resonant with the site resonators. 

In the notation employed in the Main Text, the field amplitudes $a^{(1)}_{1}\simeq a^{(1)}_{2}$ and $a^{(2)}$ would be labelled $a^{(1,1)}_{-}$ and $a^{(1,2)}_{-}$, respectively. Similar equations can be obtained for the field amplitudes of the $+$ pseudospin. 
In the Main Text we have taken for simplicity $k_{\ell}=k_{\rm w}$ and $t_{\ell}=t_{\rm w}$. 
It is straightforward to generalize these equations to account for the position shift of the link resonators, and to include the couplings with the other resonators of the lattice, which give rise to analogous coupling and losses terms as those appearing in Eqs.~\eqref{eq:CoupledModeRight} and~\eqref{eq:CoupledModeLeft}. 

An S-shaped element embedded by the site ring resonators with length $L_{\rm s}$ provides an additional contribution $\gamma_{\rm s}$ to the losses and, on top of this, it adds a term in the $-$ pseudospin equations of the form
\begin{align}
    i\dot{a}_-=\ldots -i\frac{c}{n_{\rm L}L_{\circ}}2k^2_{\rm s}e^{i\frac{\omega}{c}n_{\rm L}\frac{L_{\circ}}{4} }e^{i\frac{\omega}{c}n_{\rm L} L_{\rm s} } a_{+}\,.
\end{align}
Here, the prefactor can be understood by considering the length of the optical path connecting the points of the resonator in which the $a_{\pm}$ amplitudes are measured, and the fact that light couples into the S waveguide at two different positions. A complete derivation of such a contribution can be found in Ref.~\cite{MunozDeLasHeras_2021b}.

\setcounter{figure}{0}
\renewcommand{\thefigure}{B\arabic{figure}}

\begin{figure}[htpb]
    \centering
    \includegraphics[width=0.5\textwidth]{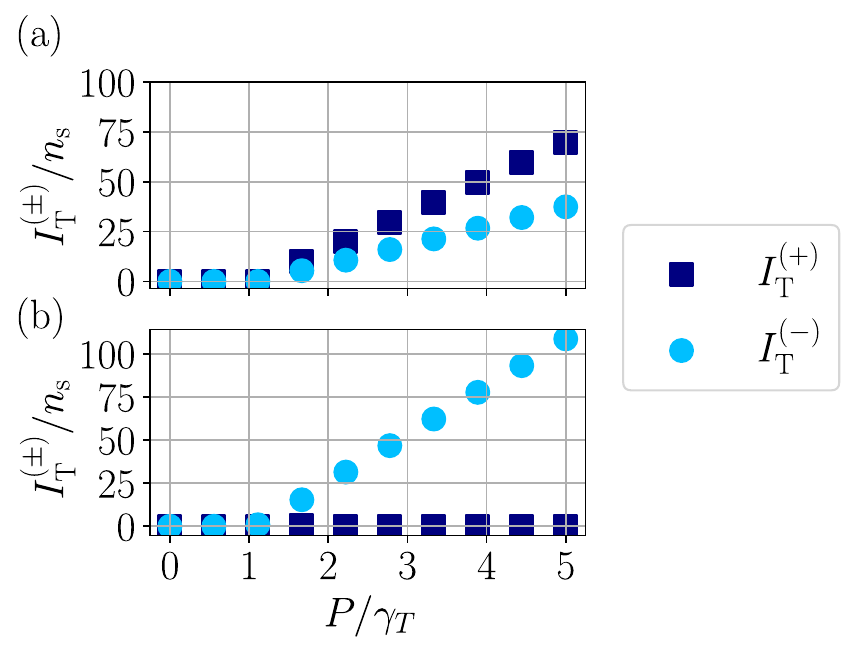}
    \caption{Total intensity $I^{(\pm)}_{\rm T}$ in the $+$ and $-$ pseudospins inside the photonic QSH lattice 
    as a function of the  pump rate $P$, in units of the total loss rate $\gamma_{\rm T}$. The edge resonators of the array feature a non-local saturable gain ($g=1$). We do not consider neither backscattering (i.e., $\beta^{\rm (BS)}_{\pm,\mp}=0$) nor Kerr non-linearity (i.e., $n_{\rm NL}=0$). (a) Ring resonator lattice ($\gamma_{\rm s}=0$) where the intensity can be arbitrarily distributed between the two pseudospin states. (b) TJR lattice with an S-waveguide coupling loss rate $\gamma_{\rm s}=0.20J$.
    }
    \label{fig:active g=1 I vs P}
\end{figure}

\section{Non-local optical non-linearities}
\label{app:nonlocal nonlinearity gain}

\begin{figure*}[htbp]
    \centering
    \includegraphics[width=0.95\textwidth]{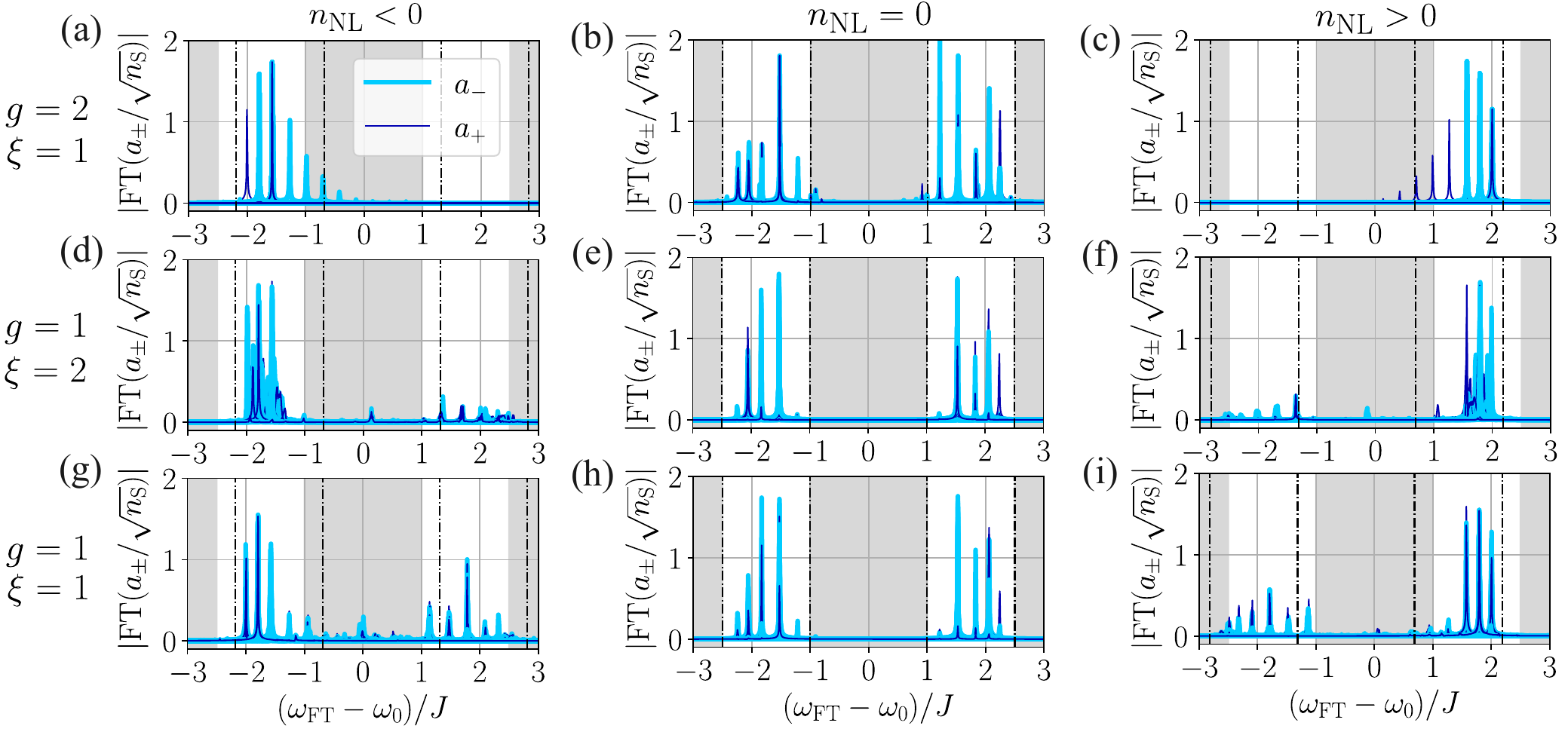}
    \caption{Emission spectrum in a lasing QSH array of ring resonators. All panels show the Fourier transform of the field amplitudes $|$FT$(a_{\pm})|$ in the top left (1,1) resonator 
    as a function of the Fourier transform frequency $\omega_{\rm FT}$. 
    Dark (light) blue solid lines correspond to the $+$ ($-$) pseudospin. The pump rate is fixed to $P=4\gamma_{\rm T}$. 
    No backscattering is considered (i.e., $\beta^{\rm (BS)}_{\pm,\mp}=0$). Each panel shows results calculated using 40 realizations. The shaded regions represent the bulk modes of the QSH array in the linear regime. In all panels, dashed-dotted lines represent the shift of the topological bandgaps due to the Kerr non-linearity. Panels (a,d,g) correspond to a non-linear refractive index $n_{\rm NL}=-10^{-5}n^{-1}_{\rm s}$. Panels (b,e,h) are computed with $n_{\rm NL}=0$. Panels (c,f,i) are calculated using $n_{\rm NL}= 10^{-5}n^{-1}_{\rm s}$. (a-c) Local saturable gain ($g=2$) and non-local Kerr non-linearity ($\xi=1$). (d-f) Non-local saturable gain ($g=1$) and local Kerr non-linearity ($\xi=2$). (g-i) Non-local saturable gain ($g=1$) and non-local Kerr non-linearity ($\xi=1$).}
    \label{fig:active nonlocal fourier}
\end{figure*}
\begin{figure*}[htbp]
    \centering
    \includegraphics[width=0.95\textwidth]{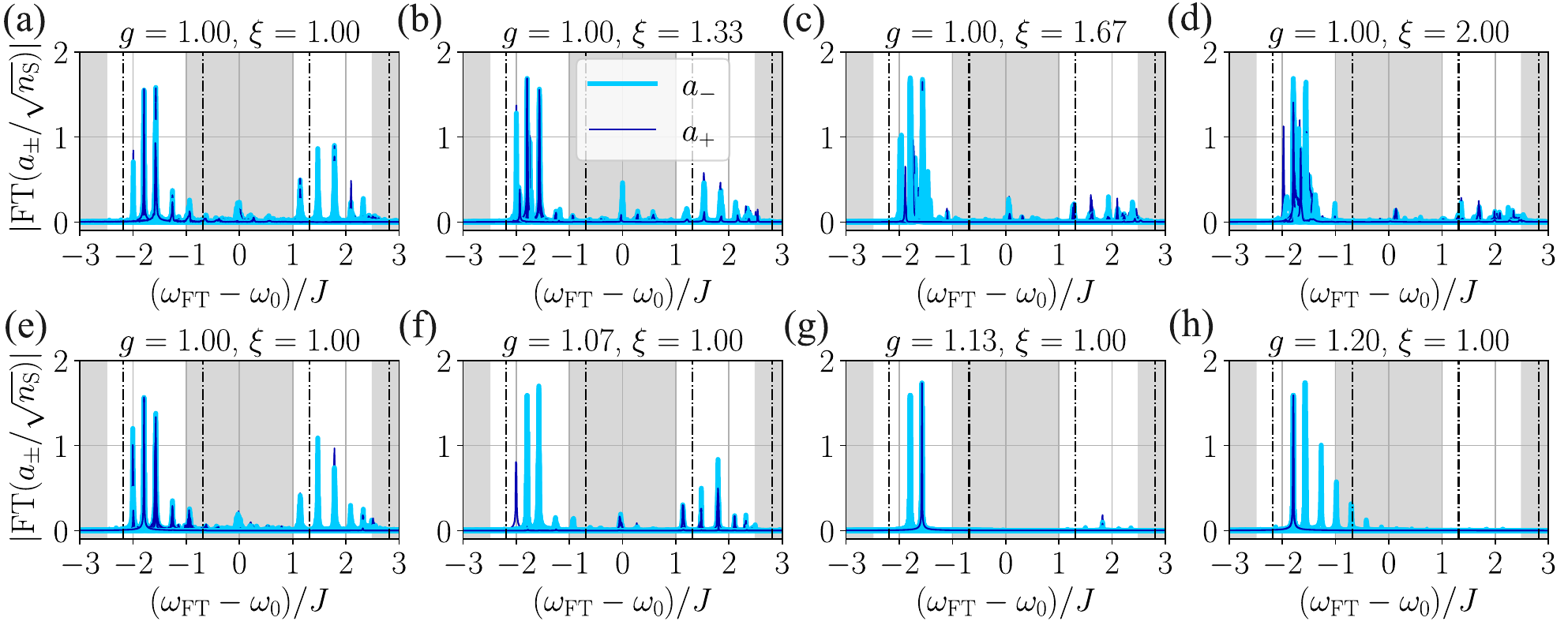}
    \caption{Emission spectrum in a lasing QSH array of ring resonators. All panels show the Fourier transform of the field amplitudes $|$FT$(a_{\pm})|$ in the top left (1,1) resonator 
    as a function of the Fourier transform frequency $\omega_{\rm FT}$. 
    Dark (light) blue solid lines correspond to the $+$ ($-$) pseudospin. The pump rate is fixed to $P=4\gamma_{\rm T}$. 
    No backscattering is considered (i.e., $\beta^{\rm (BS)}_{\pm,\mp}=0$). Each panel shows results calculated using 40 realizations. In all panels, we consider $n_{\rm NL}=-10^{-5}n^{-1}_{\rm s}$. The shaded regions represent the bulk modes of the QSH array in the linear regime. In all panels, dashed-dotted lines represent the shift of the topological bandgaps due to the Kerr non-linearity. Panels (a-d) are calculated for $g=1$ and several values of $\xi$ ranging between $1$ and $2$. Panels (e-h) are calculated for $\xi=1$ and several values of $g$ ranging between $1$ and $1.2$.
    }
    \label{fig:active nonlocal fourier g=xi=1}
\end{figure*}

In this Appendix we explore the impact of a non-local optical nature of the Kerr-type optical non-linearities and of the saturable gain in our model of the QSH topological insulator laser.

Let us start by considering a local Kerr non-linearity, as studied in the Main Text. In this case, the spatial dependence of the non-linear term has the form
%
%
\begin{equation}
    |a_+(x)+a_-(x)|^2 (a_+(x)+a_-(x))\,,
\end{equation}
as a function of the position $x$ inside a site resonator, where $a_\pm(x)\propto e^{\pm i k x}$ are plane waves propagating in opposite directions. Expanding the product and isolating the terms that go as $a_\pm(x)$, one finds that each pseudospin $\pm$ acquires a term of the form 
\begin{align}
    |a_\pm|^2 a_\pm + 2|a_\mp|^2 a_\pm,
\end{align}
which is the one present in Eq.~\eqref{eq:coupled mode}.

However, if the optical non-linearity is non-local, we can write
\begin{align}
    &\left[\int|a_+(x')+a_-(x')|^2 dx'\right]a_\pm(x)=
    \nonumber\\
    &= \int\left[
    |a_+(x')|^2 + |a_-(x')|^2 + 2\textrm{Re}[a^*_+(x')\,a_-(x')]
    \right]\,dx'
    \nonumber\\
    &\times a_\pm (x)
= [|a_+|^2 +|a_-|^2] \, a_\pm
\end{align}
without the bosonic factor $2$. 

From this calculation, we conclude that the Kerr non-linearity terms in the coupled-mode equations~\eqref{eq:coupled mode} for each site read
\begin{align}
    |a_\pm|^2 a_\pm + \xi|a_\mp|^2 a_\pm,
\end{align}
with $\xi=1$ (2) for a non-local (local) Kerr non-linearity.

Similarly, for the saturable gain triggering lasing one can carry an analog derivation to show that
\begin{align}
    i\frac{P}{1+\frac{1}{n_{\rm s}}
    \left[|a_\pm|^2 + g |a_\mp|^2 \right]}a_\pm
\end{align}
is the term to be included in the coupled-mode Eq.~\eqref{eq:coupled mode} for each site when one wants to account for a non-local ($g=1$) or local ($g=2$) saturable gain.
The results shown in Secs.~\ref{sec:passive} and~\ref{sec:active} of the Main Text were calculated using local optical non-linearities (i.e., with $\xi=2$ and $g=2$). However, in this Appendix we study the effect of non-local non-linearities.

\begin{figure}[!htbp]
    \centering
    \includegraphics[width=0.5\textwidth]{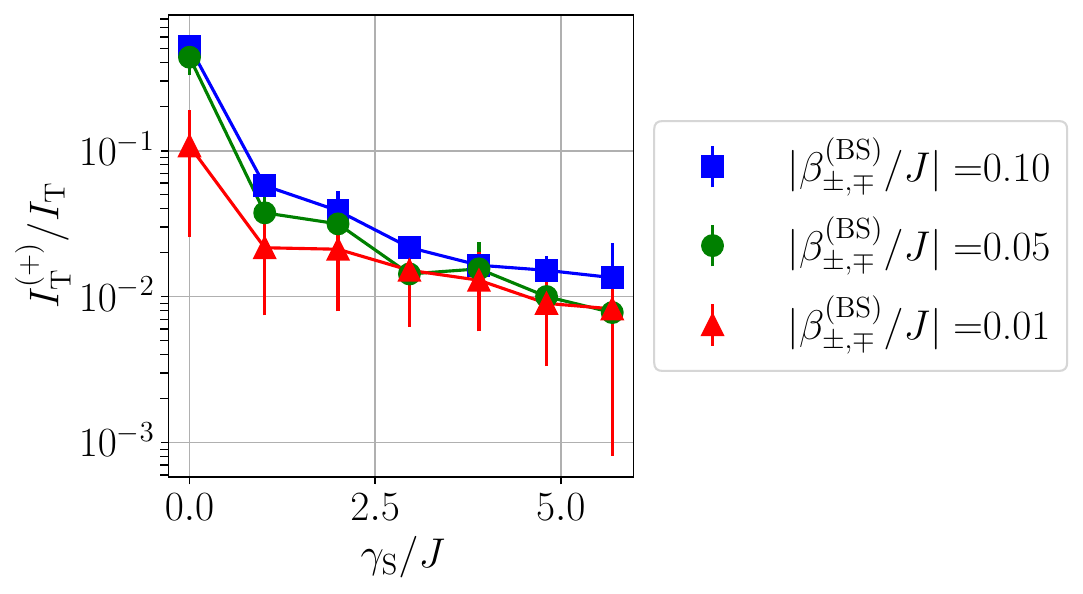}
    \caption{Active QSH array of TJRs with Hermitian backscattering and a non-local saturable gain ($g=1$). We employ a local Kerr non-linearity ($\xi=2$) of strength $n_{\rm NL}= 10^{-5}n^{-1}_{\rm s}$, and we set the pump rate to $P=4\gamma_{\rm T}$. 
    The TJRs enforce lasing in the $-$ pseudospin. We plot the total intensity in the $+$ pseudospin $I^{(+)}_{\rm T}$ in units of the total intensity in the two pseudospins $I_{\rm T}$ as a function of the S-waveguide coupling loss rate $\gamma_{\rm s} = ck^2_{\rm s} / L_\circ n_{\rm L}$ for three values of the backscattering modulus $|\beta^{\rm (BS)}_{\pm,\mp}|$ (in units of the tunneling rate $J$). Squares, circles, and triangles are calculated as the average over 10 realizations of the system. In each realization the backscattering phases and the initial values of the field amplitudes are randomly chosen at each site. Error bars represent the standard deviation. Lines are added as a guide to the eye.}
    \label{fig:active nonlocal backscattering}
\end{figure}

We start by considering an array with a non-local saturable gain in the edge resonators (i.e., $g=1$), no backscattering (i.e., $\beta^{\rm (BS)}_{\pm,\mp}=0$), and no Kerr non-linearity (i.e., $n_{\rm NL}=0$). In Fig.~\ref{fig:active g=1 I vs P} we plot the total intensity $I^{(\pm)}_{\rm T}$ in the $+$ and $-$ pseudospins inside the photonic QSH array as a function of the pump rate $P$. Each panel shows a single realization of the system, the two starting from random initial conditions for the field amplitudes. Panel (a) shows the results for a ring resonator array (i.e., with $\gamma_{\rm s}=0$), while panel (b) displays those for a TJR array with an S-waveguide coupling loss rate $\gamma_{\rm s}=0.20J$. In the former case, the symmetry between the two pseudospins in the coupled-mode equations when $g=1$ allows the system to simultaneously lase in the two whispering-gallery modes of the TJRs with an arbitrary power distribution between the two. This type of solution is impossible when a local saturable gain is considered, as the $g=2$ factor breaks the symmetry between the two pseudospins and only allows lasing in one pseudospin or with equal intensities in the two pseudospins. When the S-shaped element is added [panel (b)], lasing can only take place in the $-$ pseudospin.

As a next step, we check whether the non-local optical non-linearities affect the frequency spectrum of lasing. To do it, in Fig.~\ref{fig:active nonlocal fourier} we compute 40 realizations (each starting from random initial conditions for the field amplitudes) of the QSH laser featuring ring resonators ($\gamma_{\rm s}=0$), and we plot the Fourier transform of the field amplitudes $|$FT$(a_\pm)|$ in the $+$ and $-$ pseudospins of the top left (1,1) resonator as a function of the frequency $\omega_{\rm FT}$. In all panels, we fix the pump rate to $P=4\gamma_{\rm T}$, and no backscattering is considered ($\beta^{\rm (BS)}_{\pm,\mp}=0$). We consider the three cases that were not studied in the Main Text of this article: a local saturable gain and non-local Kerr non-linearity [$g=2$ and $\xi=1$, plotted in panels (a-c)], a non-local saturable gain and local Kerr non-linearity [$g=1$ and $\xi=2$, plotted in panels (d-f)], as well as a non-local saturable gain and non-local Kerr non-linearity [$g=1$ and $\xi=1$, plotted in panels (g-i)]. 

In the first situation ($g=2$ and $\xi=1$) the behavior is substantially the same as the one found for a local saturable gain and a local Kerr non-linearity: at each realization, lasing occurs either in a single pseudospin state or with equal intensities in the two pseudospins; negative values of the non-linear refractive index $n_{\rm NL}$ lead to lasing solely in the bottom topological gap, while positive ones trigger lasing only in the upper topological gap. For $n_{\rm NL}=0$, lasing is possible in the two topological gaps. 
In the second case ($g=1$ and $\xi=2$) each realization can feature simultaneous lasing in the two pseudospins, as the $g=1$ factor associated to the non-local character of the saturable gain does not break the symmetry between the two pseudospins. However, as in the previous case, lasing still occurs in a single topological gap determined by the sign of the Kerr non-linearity.
Also in the third case ($g=\xi=1$), each realization may lase simultaneously in the two pseudospins. In contrast to the previous two cases, for a finite Kerr non-linearity some realizations may exhibit simultaneous lasing at frequencies in the two topological gaps [see panels (g) and (i)]. However, even in these realizations the amplitude of the Fourier components $|$FT$(a_\pm)|$ is larger in the topological gap stable for the other combinations of $g$ and $\xi$ studied, i.e., the bottom one for for $n_{\rm NL}<0$, and the top one for $n_{\rm NL}>0$.

To shine more light on this behavior, in Fig.~\ref{fig:active nonlocal fourier g=xi=1} we perform an analog analysis to that of Fig.~\ref{fig:active nonlocal fourier}, but focusing on values of $g$ and $\xi$ close to $1$. We take $P=4\gamma_{\rm T}$ and $n_{\rm NL}=-10^{-5}n^{-1}_{\rm s}$. In panels (a-d), we fix $g=1$ and we calculate the Fourier transform amplitudes $|$FT$(a_\pm)|$ as a function of $\omega_{\rm FT}$ for different values of $\xi$ ranging from $\xi=1$ to $\xi=2$. As $\xi$ increases, the amplitude of the Fourier components in the upper topological gap decreases. Already for $\xi=1.67$ [panel (c)] the values of $|$FT$(a_\pm)|$ in the upper gap are substantially smaller than the values of $|$FT$(a_\pm)|$ found in bottom one. For $\xi=2$ [panel (d)], the system only features sizable Fourier components in the bottom topological gap, as expected. The decrease of the lasing components in the upper topological gap is even more dramatic when we fix $\xi=1$ and we increase $g$ [panels (e-h)]. Already for $g=1.13$, the Fourier components in the upper topological gap are practically extinct, and for $g=1.2$, we recover almost perfect single-gap lasing in the bottom topological gap.

\setcounter{figure}{0}
\renewcommand{\thefigure}{C\arabic{figure}}
\begin{figure*}[!htbp]
    \centering
    \includegraphics[width=\textwidth]{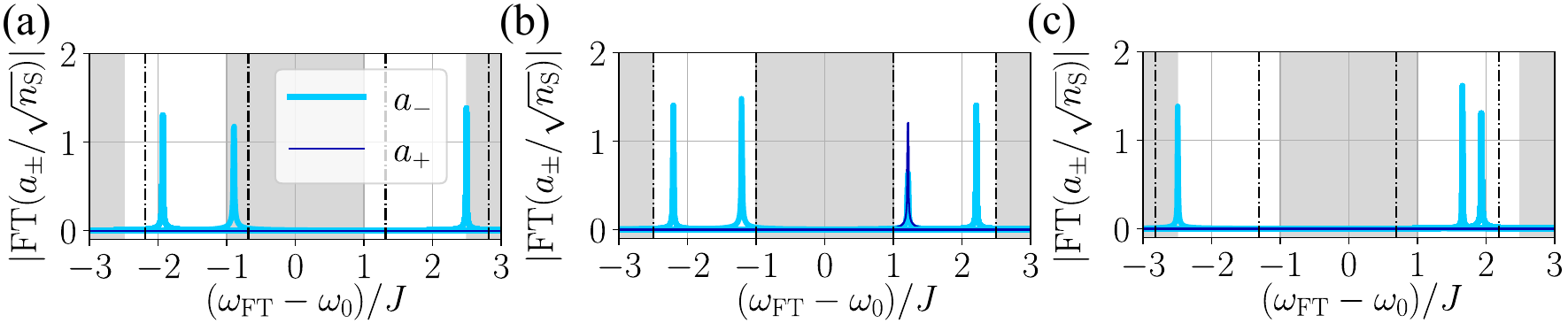}
    \caption{
    Frequency-dependent spectrum $|$FT$(a_{\pm})|$ of the field amplitudes $a_{\pm}$ 
    in the top-left (1,1) resonator of an active QSH laser of TJRs with $\gamma_{\rm s}=0.2J$, as a function of the Fourier transform frequency $\omega_{\rm FT}$. 
    The pump rate is fixed to $P=4\gamma_{\rm T}$. No backscattering is considered (i.e., $\beta^{\rm (BS)}_{\rm \pm,\mp}=0$). Data is calculated using 40 realizations, each starting from random initial conditions.  
    Dark (light) blue solid lines correspond to the $+$ ($-$) pseudospin. The shaded regions represent the bulk bands of the QSH array in the linear regime. Each panel corresponds to a different value of the Kerr non-linearity strength: (a) $n_{\rm NL}=-10^{-5}n^{-1}_{\rm s}$, (b) $n_{\rm NL}=0$, and (c) $n_{\rm NL}= 10^{-5}n^{-1}_{\rm s}$. In all panels, dashed-dotted lines represent the shift of the topological bandgaps due to the Kerr non-linearity.
    }
    \label{fig:App C 4x4 spectrum}
\end{figure*}

Finally, to test the effect of the S-shaped waveguide in the presence of Hermitian backscattering, we set the Kerr non-linearity to $n_{\rm NL}= 10^{-5}n^{-1}_{\rm s}$, and perform the same analysis done in Fig.~\ref{fig:active BS}, but in this case for a non-local saturable gain ($g=1$) and a local Kerr non-linearity ($\xi=2$). We remind the reader that the TJRs couple the $+$ into the $-$ pseudospin, thus triggering lasing in the CCW- helical mode for the value of $n_{\rm NL}$ chosen here. The results, calculated for a fixed pump rate $P=4\gamma_{\rm T}$, are shown in Fig.~\ref{fig:active nonlocal backscattering}, where we plot the total intensity in the $+$ pseudospin inside the photonic QSH lattice $I^{(+)}_{\rm T}$ as a function of the S-waveguide coupling loss rate $\gamma_{\rm s}$, for different values of the backscattering modulus $|\beta^{\rm (BS)}_{\pm,\mp}|$. Each point is calculated using 10 realizations of the QSH laser, each starting from random initial conditions and featuring random, independent phases for the backscattering at each site resonator. 

Although we get a similar behavior to that observed in Fig.~\ref{fig:active BS}, with $I^{(+)}_{\rm T}$ decreasing with $\gamma_{\rm s}$ for all values of $|\beta^{\rm (BS)}_{\pm,\mp}|$ employed, a key difference is that in the non-local saturable gain ($g=1$) case the S-shaped element is not as effective as in the local ($g=2$) one. For instance, for $g=1$ and $\gamma_{\rm s}=5.7J$ the intensity in the $+$ pseudospin takes a value around $10^{-2}I_{\rm T}$ for a backscattering rate $|\beta^{\rm (BS)}_{\pm,\mp}|=0.05J$, while for $g=2$ we obtained $I^{(+)}_{\rm T}\simeq  3\times 10^{-4}I_{\rm T}$ for the same backscattering rate. This is due to the asymmetry in the saturable gain term introduced in the non-local ($g=2$) case, which, in agreement with Eq.~(23) of Ref.~\cite{MunozDeLasHeras_2021b}, reinforces the action of the S-shaped waveguide and thus further suppresses lasing in the unfavored pseudospin.

\begin{figure}[!htbp]
    \centering
    \includegraphics[width=0.5\textwidth]{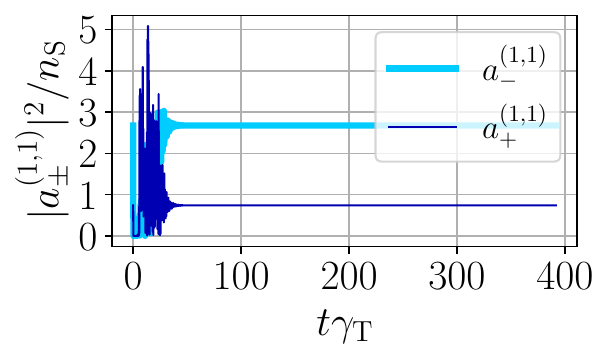}
    \caption{
    Intensity $|a^{(1,1)}_\pm|^2$ in the top left $(1,1)$ resonator in a single realization of an active $4\times4$ QSH array of TJRs as a function of time $t$. The QSH array is formed by TJRs coupling the $+$ into the $-$ pseudospin with $\gamma_{\rm s}=0.2J$. The pump rate is $P=4\gamma_{\rm T}$, the non-linear refractive index is set to $n_{\rm NL}=0$, and no backscattering is present (i.e., $\beta^{\rm (BS)}_{\pm,\mp}=0$).
    }
    \label{fig:App C strong CW}
\end{figure}
%

\section{Effect of system size on topological lasing}
\label{app:effect system size}

In this Appendix we explore how the size of the QSH array affects the single-mode topological lasing shown for an $8\times 8$ lattice in Fig.~\ref{fig:active fourier spectrum}. To do it, we study the emission spectrum of a smaller, $4\times 4$ system. For such a small array size, we expect finite-size effects to play an important role in lasing. 

Fig.~\ref{fig:App C 4x4 spectrum} shows the analog of Fig.~\ref{fig:active fourier spectrum} for a $4\times 4$ lattice: we plot the Fourier transform of the field amplitudes $|$FT$(a_{\pm})|$ in the top-left (1,1) resonator for 40 realizations of a QSH laser. Each realization is computed independently, considering random values of the initial field amplitudes. The QSH array considered here is formed by TJRs with and S-shaped element coupling the $+$ into the $-$ pseudospin and featuring $\gamma_{\rm s}=0.2J$. No backscattering is considered, i.e., we take $\beta^{\rm (BS)}_{\pm,\mp}=0$. To address the role of the Kerr non-linearity, we consider three values of the non-linear refractive index.
For $n_{\rm NL}=- 10^{-5}n^{-1}_{\rm s}$ [panel (a)], lasing is observed for the $-$ pseudospin only at several frequencies inside both Kerr-shifted topological gaps. This contrasts with the $8\times8$ case, where the Kerr non-linearity rendered lasing in the upper topological gap dynamically unstable~\cite{Loirette2021}. An analog situation is observed for $n_{\rm NL}= 10^{-5}n^{-1}_{\rm s}$ [panel (c)]: while for the $8\times8$ lasing was only possible in the upper topological gap, for a $4\times4$ system lasing is observed inside the two topological gaps. 
These results can be understood by noting that the edge of a $4\times 4$ lattice contains only a few sites, so the corresponding edge‐mode dispersion is sampled at a small set of discrete momenta. As a consequence, the edge states dispersion is not fully resolved and departs significantly from the ideal continuous limit. Thus, both topological gaps can support stable lasing even in the presence of a finite Kerr non-linearity. This behavior contrasts with the larger $8\times8$ system discussed in the Main Text, where the longer perimeter hosts a denser set of edge modes, the edge dispersions are more finely resolved, and the Kerr non-linearity selectively destabilizes one of the two topological gaps, giving rise to the single-mode lasing reported in Fig.~\ref{fig:active fourier spectrum}. 

Finally, in the absence of Kerr non-linearity, i.e., for $n_{\rm NL}=0$ [panel (b)], lasing is also observed in the two topological gaps. However, in the $4\times4$ array approximately $10\%$ of the computed realizations exhibit finite lasing in the $+$ pseudospin, although weaker than in the $-$ pseudospin. An example of the convergence of one of such finite CW lasing realizations in shown in Fig.~\ref{fig:App C strong CW}. This is in contrast to what was found for an $8\times8$ lattice, where the S-shaped element inside TRJs granted single-pseudospin lasing for all realizations.
We attribute the appearance of a small number of realizations converging to steady states with weak but finite lasing in the $+$ pseudospin to finite-size effects that weaken the directional asymmetry introduced by the TJRs. In small systems, hybridization between edge states and reduced mode competition due to the smaller number of edge states available can diminish the pseudospin selectivity. As a result, the $+$ pseudospin is not fully suppressed. We also note that the realizations exhibiting finite lasing in the $+$ pseudospin disappear when a finite Kerr non-linearity is introduced. This can be understood by noting that the Kerr-induced frequency shifts act as a non-linear detuning mechanism, which penalizes the weaker pseudospin more strongly [see Eq.~\eqref{eq:coupled mode}]. Thus, we expect this type of realizations to be absent in real systems with sizable values of the Kerr non-linearity.
Note that for a single-resonator TJR laser with the S-shaped waveguide coupling the $+$ into the $-$ mode similar metastable solutions exist in which, for small values of $\gamma_{\rm s}$, the $+$ mode is populated with a finite intensity, as was reported in Ref.~\cite{MunozDeLasHeras_2021b}. However, in that case the solutions satisfy $|a_+|^2 > |a_-|^2$.

\section{Dynamical stability of QSH topological lasing}
\label{app:dynamical stability}

In this Appendix we analyze the dynamical stability of the steady--state
solutions of Eq.~\eqref{eq:coupled mode}. We write the field amplitudes as
$\tilde a_{\pm}^{(n_x,n_y)}(t)
=\tilde a_{\pm,{\rm ss}}^{(n_x,n_y)}+\delta\tilde a_{\pm}^{(n_x,n_y)}(t)$,
where the steady state $\tilde a_{\pm,{\rm ss}}^{(n_x,n_y)}$ satisfies
$d\tilde a_{\pm,{\rm ss}}^{(n_x,n_y)}/dt=0$, and
$|\delta\tilde a_{\pm}^{(n_x,n_y)}|\ll 
|\tilde a_{\pm,{\rm ss}}^{(n_x,n_y)}|$ denote small fluctuations.
Keeping only terms linear in $\delta\tilde a_{\pm}^{(n_x,n_y)}$, the equations of motion take
the Bogoliubov form
\begin{equation}
i\frac{d}{dt}\,\boldsymbol{\delta a}
= A\,\boldsymbol{\delta a},
\label{eq:FluctuationDynamics}
\end{equation}
where the fluctuation vector is
\begin{align}
\boldsymbol{\delta a}
=&\bigl(
\delta\tilde a^{(1,1)}_{+},
\delta\tilde a^{(1,1)*}_{+},
\delta\tilde a^{(1,1)}_{-},
\delta\tilde a^{(1,1)*}_{-},
\dots,
\nonumber\\&
\delta\tilde a^{(N_x,N_y)}_{+},
\delta\tilde a^{(N_x,N_y)*}_{+},
\delta\tilde a^{(N_x,N_y)}_{-},
\delta\tilde a^{(N_x,N_y)*}_{-}
\bigr)^{T},
\label{eq:delta_vector_def}
\end{align}
\setcounter{figure}{0}
\renewcommand{\thefigure}{D\arabic{figure}}
\begin{figure*}[!htbp]
    \centering
    \includegraphics[width=\textwidth]{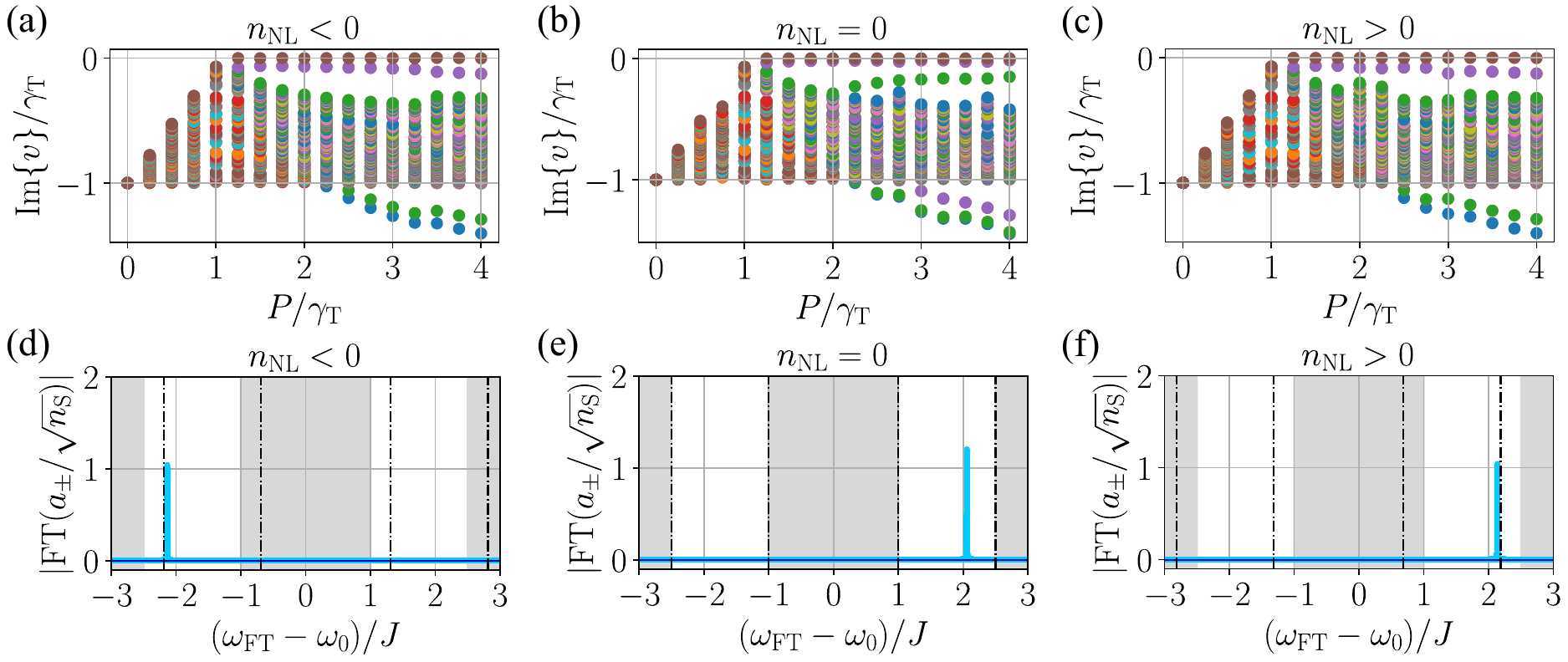}
    \caption{
    (a,b,c) Imaginary part of the eigenvalues Im$\{v\}$ of the dynamical stability matrix $A$ in Eq.~\eqref{eq:FluctuationDynamics} as a function of the pump rate $P$ for an $8\times 8$ QSH array of TJRs with $\gamma_{\rm s}=0.2J$. Both $v$ and $P$ are normalized in units of the total loss rate $\gamma_{\rm T}$. For every value of $P$, each panel shows a single stochastic realization of the QSH topological laser for different Kerr non-linearity strengths: (a) $n_{\rm NL}=-10^{-5}n_{\rm s}^{-1}$, (b) $n_{\rm NL}=0$, and (c) $n_{\rm NL}=10^{-5}n_{\rm s}^{-1}$. Backscattering is not considered (i.e., $\beta^{\rm (BS)}_{\pm,\mp}=0$). 
    (d,e,f) Frequency-dependent spectrum $|$FT$(a_{\pm})|$ of the field amplitudes $a_{\pm}$ in the top-left resonator $(1,1)$ of the active QSH lattice of TJRs, computed for a pump rate $P=4\gamma_{\rm T}$, for the same realizations whose stability spectra appear in panels (a,b,c), respectively. Dark (light) blue curves correspond to the $+$ ($-$) pseudospin. Shaded regions indicate the linear regime bulk bands of the QSH array, while dashed-dotted vertical lines represent the shift of the topological bandgaps due to the Kerr non-linearity.
    }
    \label{fig:App D}
\end{figure*}
and the matrix $A$ is the $4N_xN_y\times4N_xN_y$ block matrix
\begin{widetext}
    \begin{equation}
    A=\begin{pmatrix}
    A_{(1,1),(1,1)} & A_{(1,1),(1,2)} & \cdots & A_{(1,1),(N_x,N_y)}\\[2mm]
    A_{(1,2),(1,1)} & A_{(1,2),(1,2)} & \cdots & A_{(1,2),(N_x,N_y)}\\[1mm]
    \vdots & \vdots & \ddots & \vdots\\[1mm]
    A_{(N_x,N_y),(1,1)} & A_{(N_x,N_y),(1,2)} &
    \cdots & A_{(N_x,N_y),(N_x,N_y)}
    \end{pmatrix},
    \label{eq:A_block_form}
    \end{equation}
\end{widetext}
with each
site $(n_x,n_y)$ contributing a $4\times4$ on-diagonal block $A_{(n_x,n_y),(n_x,n_y)}$ [given by Eq.~(4) of Ref.~\cite{MunozDeLasHeras_2021b}, which was derived to study single-resonator lasing], and nearest-neighbour couplings
entering as off-diagonal $4\times4$ blocks $A_{(n_x,n_y),(n'_x,n'_y)}$ with $n_x\neq n'_x$ and $n_y\neq n'_y$. The off-diagonal blocks take the form
\begin{widetext}
    \begin{equation}
    A_{(n_x,n_y),(n'_x,n'_y)} =
    \begin{pmatrix}
    J^{(+)}_{n_x,n_y\to n'_x,n'_y} & 0 & 0 & 0 \\[2mm]
    0 & J^{(+)\,*}_{n_x,n_y\to n'_x,n'_y} & 0 & 0 \\[2mm]
    0 & 0 & J^{(-)}_{n_x,n_y\to n'_x,n'_y} & 0 \\[2mm]
    0 & 0 & 0 & J^{(-)\,*}_{n_x,n_y\to n'_x,n'_y}
    \end{pmatrix},
    \end{equation}
\end{widetext}
where the coupling matrix elements
$J^{(\pm)}_{n_x,n_y\to n'_x,n'_y}$ are given by
\begin{widetext}
    \begin{equation}
    J^{(\pm)}_{n_x,n_y\to n'_x,n'_y}
    =
    \begin{cases}
    J 
    & \text{if } n_x = n'_x \text{ and } n_y = n'_y + 1
    , \\[2mm]
    J 
    & \text{if } n_x = n'_x \text{ and } n_y = n'_y - 1
    , \\[2mm]
    J\,e^{\mp i 2\pi\alpha (N_y - n_y)}
    & \text{if } n_x = n'_x + 1 \text{ and } n_y = n'_y
    , \\[2mm]
    J\,e^{\pm i 2\pi\alpha (N_y - n_y)}
    & \text{if } n_x = n'_x - 1 \text{ and } n_y = n'_y
    , \\[2mm]
    0 
    & \text{otherwise}.
    \end{cases}
    \label{eq:Jcases}
    \end{equation}
\end{widetext}
The dynamical stability of the steady state is determined by the eigenvalues
$v$ of $A$.  The steady state is stable if and only if
$\mathrm{Im}(v)\le0$ for all eigenvalues.  

Fig.~\ref{fig:App D}(a-c) shows the imaginary part of the eigenvalues of the dynamical stability matrix $A$ in Eq.~\eqref{eq:FluctuationDynamics} as a function of the pump rate $P$ for an $8\times 8$ QSH array of TJRs, each featuring an internal S-shaped element with $\gamma_{\rm s}=0.2J$. The three panels correspond to different values of the Kerr non-linearity [$n_{\rm NL}=-10^{-5}n^{-1}_{\rm s}$ in panel (a), $n_{\rm NL}=0$ in panel (b), and $n_{\rm NL}= 10^{-5}n^{-1}_{\rm s}$ in panel (c)]. We do not consider backscattering, i.e., we set  $\beta^{\rm (BS)}_{\rm \pm,\mp}=0$. For every value of $P$ in each panel of Fig.~\ref{fig:App D}, the eigenvalues $v$ are obtained from a single realization of the QSH topological laser, employing random initial conditions for the amplitudes of each pseudospin $a_{\pm}^{(n_x,n_y)}$. These conditions are the same as those employed to produce Fig.~\ref{fig:active fourier spectrum}(d-f), therefore allowing us to study the dynamical stability of single-mode topological lasing.
For the three values of $n_{\rm NL}$ studied, we observe two distinct regimes, as expected. Below the lasing threshold, i.e., for $P<\gamma_{\rm T}$, all eigenvalues have negative imaginary parts, indicating that the trivial vacuum is dynamically stable. Above threshold, i.e., for $P>\gamma_{\rm T}$, a single eigenvalue is pinned at $\operatorname{Im}v = 0$, signaling the emergence of the Goldstone mode associated with spontaneous $U(1)$ symmetry breaking. All remaining modes satisfy $\operatorname{Im}v < 0$, as expected for a stable steady state. 
Finally, in Fig.~\ref{fig:App D}(d-f) we plot, respectively, the frequency spectrum $|$FT$(a_{\pm})|$ of the field amplitudes $a_{\pm}$ in the top-left resonator $(1,1)$ for the realizations with a pump rate $P=4\gamma_{\rm T}$ shown in Fig.~\ref{fig:App D}(a-c). As expected from Sec.~\ref{sec:active}, we obtain single-mode lasing inside a topological gap determined by the sign of $n_{\rm NL}$. For $n_{\rm NL}=0$ [panel (e)], we find a realization in which the system lases in a single frequency located in the upper topological gap, although lasing is still possible in the lower one in this case, as shown in Fig.~\ref{fig:active fourier spectrum}(e).
Overall, these results confirm the dynamical stability of the single-mode topological lasing solutions found in Sec.~\ref{sec:active}.

\bibliography{bibliography.bib}

\end{document}